\documentstyle[12pt,prd,aps,psfig,floats]{revtex}
\input psfig
\def\lsim{\mathrel{\rlap{\lower4pt\hbox{\hskip1pt$\sim$}}
    \raise1pt\hbox{$<$}}}         
\def\gsim{\mathrel{\rlap{\lower4pt\hbox{\hskip1pt$\sim$}}
    \raise1pt\hbox{$>$}}}         
\begin{document}
\begin{titlepage}

\title{Correlations of Solar Neutrino Observables for SNO}
\author{{John N. Bahcall}\thanks{jnb@ias.edu}} 
\address{School of Natural Sciences, 
Institute for Advanced Study, Princeton, NJ 08540}
\author{Plamen I. Krastev\thanks{krastev@nucth.physics.wisc.edu}}
\address{Department of Physics, University of Wisconsin, Madison, WI 53706}
\author{Alexei Yu. Smirnov\thanks{smirnov@ictp.trieste.it}}
\address{School of Natural Sciences, Institute for Advanced Study, 
Princeton, NJ 08540\\
International Center for Theoretical Physics, 34100 Trieste, Italy}
\maketitle
\vglue.3in

\begin{abstract}
Neutrino oscillation scenarios predict correlations, and zones of
avoidance, among measurable quantities such as spectral energy
distortions, total fluxes, time dependences, and flavor content. The
comparison of observed and predicted correlations will enhance the
diagnostic power of solar neutrino experiments.  A general test of all
presently-allowed (2$\nu$) oscillation solutions is that future
measurements must yield values outside the predicted zones of
avoidance. To illustrate the discriminatory power of the simultaneous
analysis of multiple observables, we map currently allowed regions of
$\Delta m^2 - \sin^2 2\theta$ onto planes of quantities
measurable with the Sudbury Neutrino Observatory (SNO). We calculate
the correlations that are predicted by vacuum and MSW (active and
sterile) neutrino oscillation solutions that are globally consistent
with all available neutrino data. We derive approximate analytic
expressions for the dependence of individual observables and specific
correlations upon neutrino oscillations parameters. We also discuss
the prospects for identifying the correct oscillation solution using
multiple SNO observables.

\end{abstract}

\end{titlepage}

\pacs{26.62.+t, 12.15.Ff, 14.60.Pq, 96.60.Jw}

\section{Introduction}
\label{sec:introduction}
After more than three decades of study, the number of proposed
particle physics solutions to the solar neutrino problem is still
increasing with time. The currently viable solutions to the available
set of experimental data include two, three, and four neutrino oscillation
 scenarios (with vacuum and MSW oscillations among active as
well as sterile neutrinos), neutrino decay, violation of Lorentz
invariance, violation of the weak equivalence principle, and
magnetic-moment transitions. Even for the simplest case of two
neutrino oscillations, there are several isolated regions in neutrino
parameter space that are consistent with all of the published data by
the chlorine, Kamiokande, SuperKamiokande, GALLEX, and SAGE
experiments.

The existing solar neutrino data provide at most (2--3)$\sigma$
indications favoring specific solutions.  Moreover, the predicted
sizes of those neutrino conversion effects that do not depend upon the
standard solar model and that can be measured well in the Sudbury
Neutrino Observatory (SNO) \cite{sno}, are typically small: from a few
per cent to about ten per cent \cite{bks2000}.  Exceptions include the
day-night asymmetry (for limited values of the oscillation parameters)
and the double ratio of the neutral- to charged-current event rates.
We will have to be lucky for the oscillation effects to be realized
near their maximal possible values. In the largest part of the $\Delta
m^2 - \sin^2 2\theta$ parameter space, currently acceptable neutrino
oscillation scenarios predict that most of the new physics effects for
SNO will typically be less than $3$ or $4$ $\sigma$ different from the
no-oscillation predictions. And, as previous experience teaches us,
Nature seems to prefer toying with us by providing ambiguous hints.
The existence of systematic effects at the several percent level
further increases the difficulty of identifying a unique solution.

In this paper, we show how the predictions for some solar neutrino
observables are correlated, or why they are uncorrelated, in the
context of specific solutions of the solar neutrino problems. We
demonstrate that the signatures of a given neutrino solution include
not only the values of specific observables, but also the correlations
among the observables. A study of the correlations (and,  where relevant,
the lack of correlations) between the different predicted values of
neutrino observables can be used to increase our understanding of the
physical processes that are occurring.\footnote{Also, numerical codes for
calculating neutrino oscillation processes can be tested by requiring
that they yield correlations predicted by analytic arguments given in
this paper.}

In addition, there are excluded regions that we call ``zones of
avoidance.'' None of the currently favored oscillation solutions
predict that the measured neutrino observables will lie within these
regions of avoided parameter space.  We stress the diagnostic value of
testing predictions that new measurable quantities lie outside these
current zones of avoidance.

The main point of this paper is that studying the predicted
correlations and zones of avoidance among  solar neutrino observables 
can add discriminatory power to solar neutrino
experiments. Although we illustrate the methodology using SNO
variables and a particular set of allowed neutrino oscillation
parameters, the diagnostic value of the predicted correlations and 
zones of avoidance are more general than the particulars of our
illustrative calculations.

\subsection{Previous discussions of correlations}
\label{subsec:previous}

Correlations of solar neutrino observables have been discussed in
  several previous investigations. Perhaps the first such discussion
  pointed out (Ref.~\cite{BB}) the lack of consistency, if no new
  physics were involved, between the total rates observed in the
  chlorine~\cite{chlorine} and in the Kamiokande~\cite{kamiokande}
  experiments. This inconsistency was used as an argument to exclude
  astrophysical solutions of the solar neutrino problem. Kwong and
  Rosen~(Ref.~\cite{rosen}) analyzed the relations between event
  rates measured with SuperKamiokande and with the Sudbury Neutrino
  Observatory (SNO) for various MSW solutions. Even more closely
  related to what we discuss in the present paper, Folgi, Lisi, and
  Montanino \cite{bari} mapped the Large Mixing Angle (LMA) MSW and
  the Small Mixing Angle (SMA) MSW solution regions in the $\Delta m^2 -
  \sin^2 2\theta$ plane onto the plane of flux independent
  observables: the day-night asymmetry, and the shift of the first
  moment of the electron spectrum measured with SuperKamiokande and
  SNO.  The goal of the Bari group~\cite{bari} was to show how
  correlations could be used to help distinguish between the SMA and
  the LMA solutions.  The correlations of a spectrum distortion and
  the day-night asymmetry for SMA and MSW Sterile solutions were
  discussed in Ref.~\cite{AS}.

As discussed in Ref.~\cite{lma}, strong  correlation exists 
between the day-night asymmetry and the
seasonal 
variations of  signals for the MSW solutions: both effects originate 
from the same earth regeneration effect. 
For the SMA solution,  a  strong correlation exists between the total
day-night asymmetry and the event rate in the  "core bin"  
(the night bin in which neutrinos cross the core of the earth) \cite{ASmor}.
For vacuum oscillation solutions (VAC) it has  been
 pointed out in Ref.~\cite{MiSm}  that  there is  a strong correlation of
spectrum distortion and seasonal variations. 

In Ref.~\cite{bks2000}, we described the goal of analyzing
simultaneously all of the SNO observables, measured and upper limits,
in order to best constrain the allowed neutrino solutions. As a first
step in that direction, we considered some pairs of measurable
quantities but did not calculate the correlations between the
predictions in the planes formed by the observables.

In this paper, we illustrate the power of studying the correlations
between different solar neutrino observables by evaluating the
correlations between measurable quantities in the SNO experiment.  We
elucidate the physical basis for the correlations with the aid of
simple analytic approximations.

\subsection{Correlated SNO Observables}
\label{subsec:snoobservables}

We consider here the correlations between the following quantities that are measurable in the SNO experiment.

\begin{itemize}

\item
The total reduced rate of the charged-current events 
above a specified threshold:  
\begin{equation}
{\rm [CC]} ~\equiv~ \frac{N_{\rm CC}}{N_{\rm CC}^{\rm SSM}}~.
\label{rcc}
\end{equation}
Here $N_{\rm CC}$ is the observed number of events from the CC
(neutrino capture reaction by deuterium) and $N_{\rm CC}^{\rm SSM}$ is
the number expected on the basis of the BP98 standard solar
model~\cite{bp98} and no new particle physics beyond what is predicted
by the standard electroweak model.  The predictions for [CC] implied
by the six currently allowed two-neutrino oscillation solutions have
been calculated in \cite{snoshow} and are also discussed in
\cite{bks2000}. 

In what follows, we consider first the correlations of different
experimental quantities with [CC] since the CC rate is the easiest
quantity for the SNO collaboration to measure.

\item
The day-night asymmetry of the charged-current events~\cite{daynight}:  
\begin{equation}
A_{{\rm N-D}}^{\rm CC} ~\equiv~ 2\frac{{\rm
N}-{\rm D}}{{\rm N} + {\rm
D}}~.
\label{ccAdn}
\end{equation}
Here N and D are the rates of the events observed during the
night-time (N) and the day-time (D) averaged over the year (and
corrected for the seasonally changing distance between the sun and the
earth).  The contours of constant $A_{\rm N-D}^{\rm CC}$ have been
calculated in Ref.~\cite{bkdn}. In what follows, we will use the
notation $A_{\rm N-D}$ to denote the charged-current day-night effect
and will use the more cumbersome notation, $A_{\rm N-D}^{\rm CC}$,
only when there is a chance of confusion with the day-night effect
measured from neutrino-electron scattering in SuperKamiokande, $A_{\rm
N-D}^{\rm ES}$.

\item
The relative shift of the first moment of the electron  energy spectrum 
from its non-oscillation value : 
\begin{equation}
\delta T \equiv \frac{T - T_0}{T_0}~.
\label{eq:mom}
\end{equation} 
Here $T$ and $T_0$ are the first moments of the recoil electron energy
distribution calculated with and without neutrino oscillations.  The
shift has been defined in \cite{bkl-mom}; we calculated $\delta T$ in
Ref.~\cite{bks2000} for the currently allowed set of neutrino
parameters.  The distortion is expected to be smooth for all solutions
except for vacuum oscillations with large $\Delta m^2$, so that the
first moment characterizes the distortion of the recoil energy
spectrum rather well.

\item 
The double ratio of the reduced neutral-current rate (neutrino
disintegration of deuterium), $N_{\rm NC}$,
to the reduced charged-current event rate: 
\begin{equation}
\frac{\rm [NC]}{\rm [CC]} ~\equiv  ~\frac{N_{\rm NC}/N_{\rm CC}}
{(N_{\rm NC}/N_{\rm CC})^{\rm SSM}}~.
\label{nccc}
\end{equation}

We will also discuss  the ratio of the reduced
rates of neutrino-electron  
scattering [ES] and charged current events [CC]: 
[ES]/[CC], 
\begin{equation}
\frac{\rm [ES]}{\rm [CC]} ~\equiv  ~\frac{N_{\rm ES}/N_{\rm CC}}
{(N_{\rm ES}/N_{\rm CC})^{\rm SSM}}~.
\label{escc}
\end{equation}
Here [ES] $\equiv N_{\rm ES}/N_{\rm ES}^{\rm SSM}$, where $N_{\rm ES}$ is
the number of observed $\nu - e$ scattering events and $N_{\rm
ES}^{\rm SSM}$ is the number of predicted events according to the SSM.

\end{itemize}

Additional SNO observables are discussed in \cite{bks2000}. In
particular, the seasonal variations may be
significant for both vacuum\cite{gribov}
and MSW\cite{bks2000,holanda} solutions.

\subsection{Outline of this paper}
\label{subsec:outline}

In Sec.~\ref{sec:method}, we describe our method.  In the next three
sections, we study correlations related to the charged-current (CC)
events observable with SNO: $A_{\rm N-D} -$ [CC] in
Sec.~\ref{sec:r-a}, [CC] - $\delta T$ in Sec. \ref{sec:r-mom}, and
$A_{\rm N-D}-\delta T$ in Sec. \ref{sec:a-mom}.  We discuss in
Sec.~\ref{sec:nc} correlations in the plane of [NC]/[CC] and $A_{\rm
N-D}$ and in Sec.~\ref{sec:ncccdeltat} the correlations of [NC]/[CC]
and $\delta$T. In Sec.~\ref{sec:sum}, we summarize our results.  In the
Appendix, we describe the dependence on oscillation parameters of each
of the observables discussed in the main text.  Simple analytic
expressions for these dependences are presented in the Appendix; these
analytic expressions are useful for interpreting the results of
detailed numerical calculations.

\subsection{How should this paper be read? }
\label{subsec:reading}

The most efficient way to read this paper is to first obtain an
overview of what is accomplished and then to descend into the
details. We recommend that the reader begin by looking at 
Fig.~\ref{fig:cca5} to Fig.~\ref{fig:ncccm8}, which show the 
correlations and the zones of avoidance in planes constructed from
quantities that are measurable by SNO. Then we suggest that the reader
turn immediately to Sec.~\ref{sec:sum}, where we provide a succinct
summary of our principal results and conclusions. Only after having
acquired this overview, do we recommend returning to
Sec.~\ref{sec:method} in order to read about the details. 
 
\section{Method}
\label{sec:method} 

\begin{figure}[!ht]
\centerline{\psfig{figure=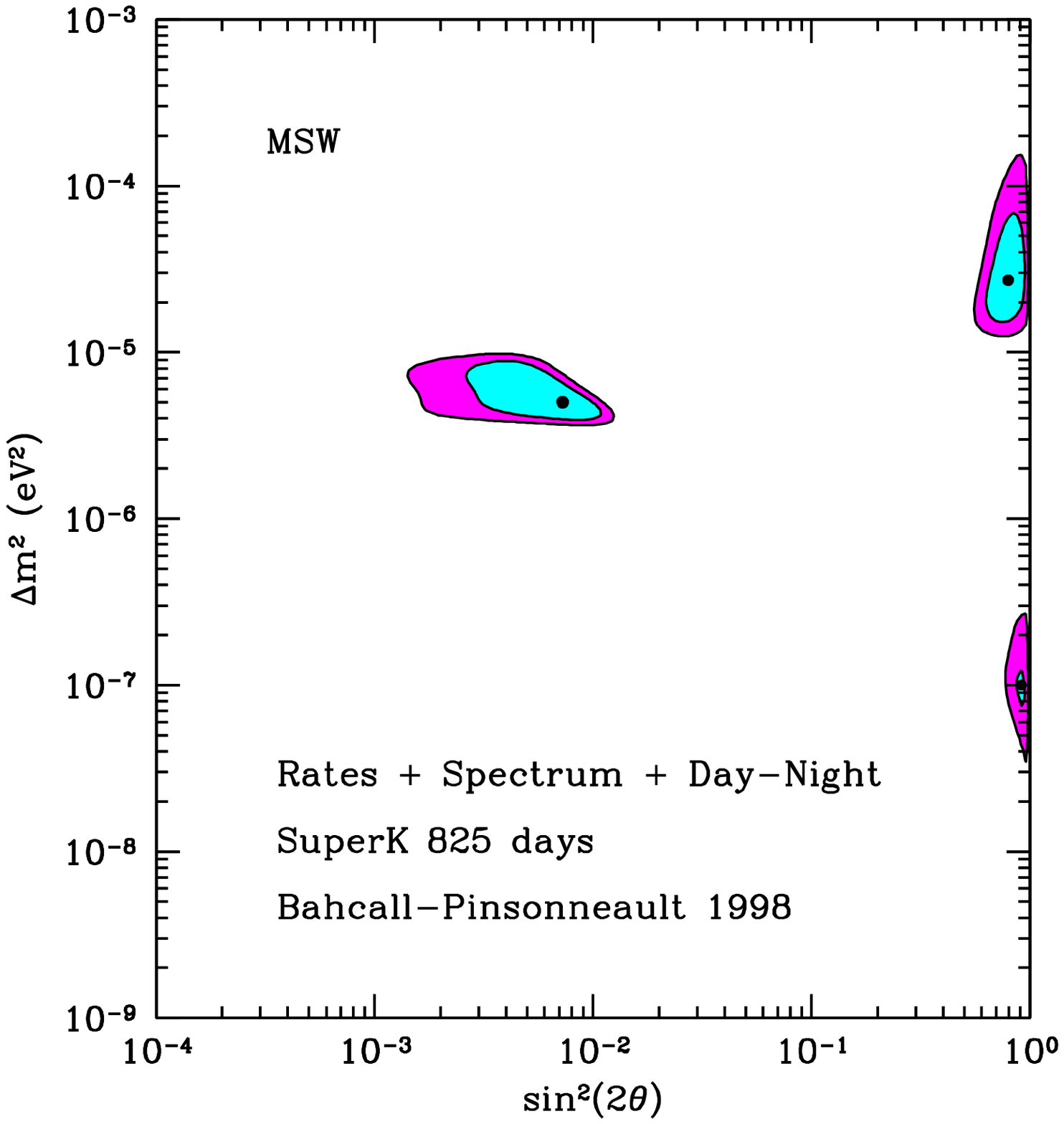,width=3.5in}}
\vglue-0.4in
\centerline{\psfig{figure=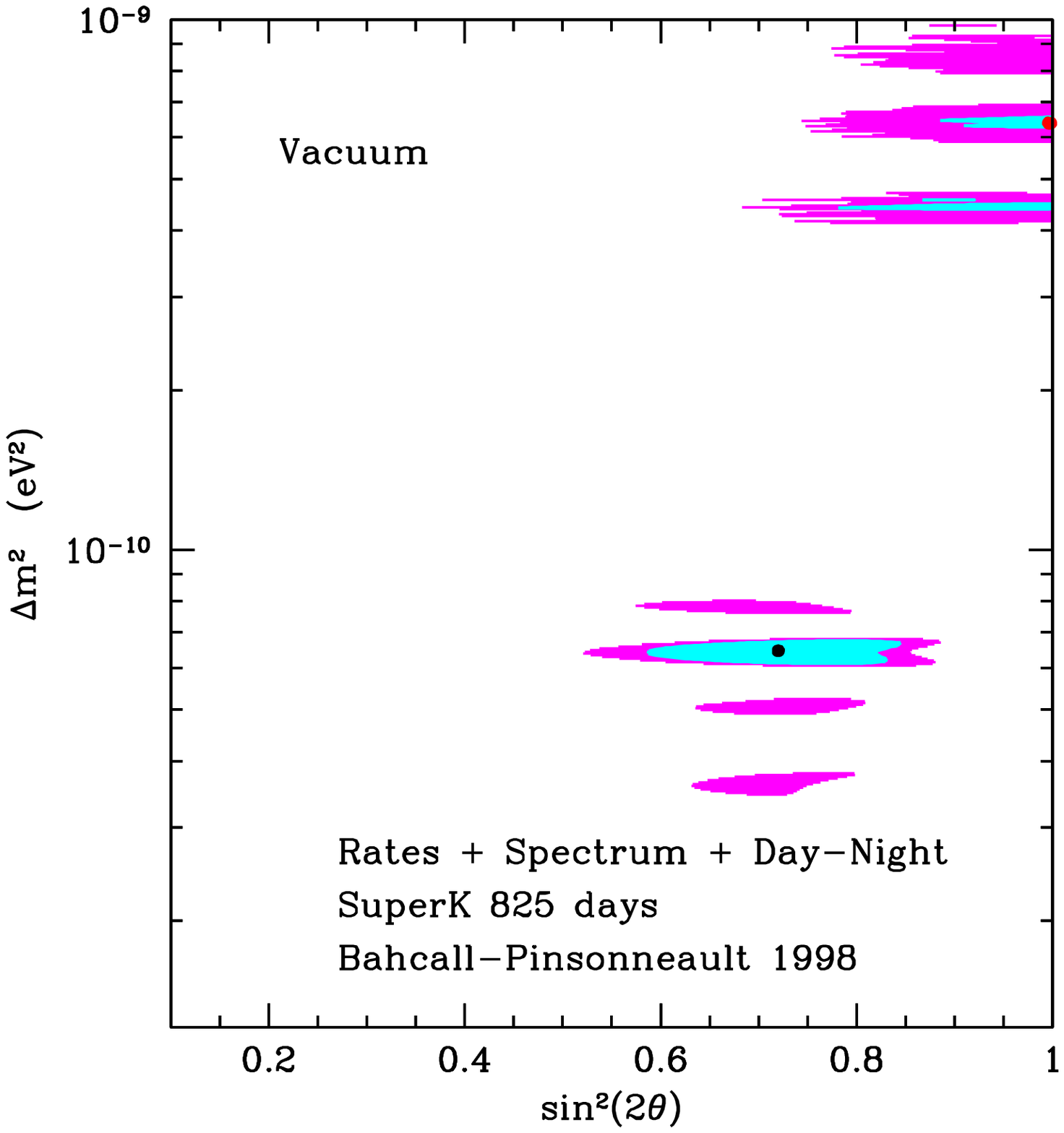,width=3.5in}}
\tightenlines
\caption[]{\small Global oscillation solutions.
The input data include the total rates in the Homestake,
Sage,  Gallex, and 
SuperKamiokande experiments, as well as the electron recoil energy
spectrum and the Day-Night effect measured by SuperKamiokande in 825
days of data taking.
Figure~\ref{fig:global}a shows the
global solutions for the allowed MSW oscillation regions,
known, respectively, as
the SMA, LMA, and LOW solutions~\cite{bks98}.
Figure~\ref{fig:global}b shows
the global solution for the allowed vacuum oscillation regions.
The CL  contours correspond, for both panels, to 
$\chi^2 = \chi^2_{\rm min} + 4.61 (9.21)$, representing  
90\% ( 99\% CL) relative to each of the best-fit solutions which are marked by
dark circles.
\label{fig:global}}
\end{figure}

For specificity, we consider correlations of observables in the SNO
experiment for the two-flavor neutrino solutions of the solar neutrino
problem. Each solution is characterized by the two oscillation
parameters, $\Delta m^2$ and $\sin^2 2\theta$.  We use the techniques
described in Ref.~\cite{bks98} to determine the allowed regions for
the oscillation parameters.  The input data used here include the
total rates in the Homestake, Sage, Gallex, and SuperKamiokande
experiments, as well as the electron recoil energy spectrum and the
Day-Night effect measured by SuperKamiokande in $825$ days of data
taking.

Fig.~\ref{fig:global} shows the acceptable regions of the solutions in
the plane of $\Delta m^2 -\sin^2 2\theta$ as determined in
Ref.~\cite{snoshow}. For our study of correlations, as
exhibited in Fig.~\ref{fig:cca5}-Fig.~\ref{fig:ncccm8}, we use the
$99$\% C.L. solutions shown in Fig.~\ref{fig:global}.

We stress that the particular topography of the predicted correlations
and the zones of avoidance depend upon the set of experimental data
that are used in finding the allowed regions and the Confidence Limit
(CL) that is adopted. We have adopted the input data and the CL
specified above.  Of course, the correlations and the zones of
avoidance will evolve as more experimental data become
available. Hopefully, the predicted correlations will become stronger
and the zones of avoidance much larger.

We perform both  numerical and semi-analytical studies
of correlations. The analytic work   provides, in many  cases, a   
simple  interpretation of  the numerical results.

For SNO, two solar neutrino fluxes are relevant: the $^8$B flux and
the $hep$ flux.  We characterize these two neutrino fluxes by the
dimensionless parameters $f_B$ and $f_{hep}$ \cite{hep,hepth}, which
are the fluxes in units of the BP98 Standard Solar Model
fluxes~\cite{bp98}.

\subsection{Mapping from neutrino space to observable space}
\label{subsec:mapping}

Predicted values of SNO  observables, 
$X$ (e.g.,  [CC], $A_{\rm N-D}$, $\delta T$, 
[NC]/[CC]),   are  functions of two 
oscillation parameters  and two  flux  parameters:  
\begin{equation}
X \equiv X(\Delta m^2, \sin^2 2\theta, f_B, f_{hep}). 
\label{xvar}
\end{equation} 
Following the same procedure as in Ref.~\cite{snoshow}, we determine
$f_B$ and $f_{hep}$ for each pair of values of the oscillation
parameters, $\Delta m^2$ and $\sin^2 2\theta$, by fitting to the total
rate and the recoil electron energy spectrum of
SuperKamiokande~\cite{superk}:
\begin{equation}
f_B = f_B (\Delta m^2, \sin^2 2\theta),~~~~ 
f_{hep}  = f_{hep}(\Delta m^2, \sin^2 2\theta). 
\end{equation}   
After this determination, the SNO observables are  functions of  two
oscillation parameters only:  
\begin{equation}
X =  X(\Delta m^2,~ \sin^2 2\theta).
\label{mapping}
\end{equation}
The correlations depend on the form of the functions,
Eq.~(\ref{mapping}); the functions are different for each solution of
the solar neutrino problem.  We give in the Appendix simple
parametrizations of the dependences for various oscillation solutions.

In what follows, we find regions in planes of ($X$, $Y$) observables
allowed by the data from all existing solar neutrino experiments.
Formally, this is equivalent to {\it mapping} the solution regions in
the $\Delta m^2 - \sin^2 2\theta$ plane onto the plane of observables
of $X$ - $Y$. For each point $\Delta m^2 - \sin^2 2\theta$ of the
solution regions, we calculate values of $X$ and $Y$.  The mapping is
given by Eq.~(\ref{mapping}).

If the region of an oscillation solution in the $\Delta m^2 - \sin^2
2\theta$ plane is projected onto a line segment or onto a narrow strip in the
$X$ and $Y$ space,  then we say that there is a strong correlation of the
observables $X$ and $Y$ for the specified solution.  In some cases,
there is a  strong
correlation only in  part of a given solution region. 

There are
various ways one might quantify the degree of correlation.  The
most appropriate way for our purpose (enhancing the
identification power of the analysis) is the following.  Let us denote by
$S_{XY}$ the area of the region in the $X - Y$ plane to which a given
solution region is projected.  Let $\Delta X$ and $\Delta Y$ be the
intervals of the observables $X$ and $Y$ in which these observables can
vary within a given solution region if we consider the variables as 
independent.  The product $\Delta X \times \Delta Y$ is the
area of the mapped region if $X$ and $Y$ are  uncorrelated.
The degree of the correlation of $X$ and $Y$ can be characterized
by the ratio:
\begin{equation}
\kappa_{XY} \equiv \frac{S_{XY}}{\Delta X \times \Delta Y}~. 
\label{eq:kappa}
\end{equation}
If the correlation parameter $\kappa_{XY} << 1$, we will say that a
strong correlation of the $X$ and $Y$ observables exists. In this
case, the allowed $X$ and $Y$ parameter space is small and ``zones of
avoidance" dominate.  For strong correlations, a combined study of the
observables $X$ and $Y$ will enhance the identification power of the
analysis.  If $\kappa_{XY} \sim 1$, there is no correlation and no
advantage to a combined study of $X$ and $Y$.

\subsection{Analytic approximations}
\label{subsec:analyticapproximations}

The accurate prediction of solar neutrino observables requires
multiple integrations over energy-dependent survival probabilities and
neutrino interaction cross sections, and also over the energy
resolution and the efficiency of detection. In spite of the
complicated nature of the accurate calculations, simple and useful
analytic results can often be found. The analytic expressions
generally contain a small number of parameters that can be determined
using the detailed numerical results. In developing analytic
approximations, we proceed as described below.

First, we determine the  functional dependence of the observables on 
the neutrino oscillation parameters, primarily through the dependence
of the survival probability, $P$, on the oscillation parameters. Thus

\begin{equation}
X(\Delta m^2, \sin^2 2\theta)  ~\sim ~ X(P(\Delta m^2, \sin^2
2\theta))~, 
\end{equation} 
where the parameter $P$ represents the survival probability after a suitable average over the energy.   
Therefore the first step is to  find expressions for observables in 
terms of $P$. 

Second, the expressions for the survival probability $P$ can usually be simplified in the restricted regions of oscillation parameters that apply to specific allowed solutions. 
 Also, averaging over relatively 
small intervals of energies (smoothing  the dependences)
often leads to further simplification.

Third, in the analytic expressions for observables, the average
neutrino energy should be taken as a fitting parameter, which is
determined by comparison of the analytic expression with the detailed
results of numerical calculations.  The energy parameter should be
fitted separately for different solutions and for different
observables. Moreover, if a given observable is described by several
terms with different dependences on energy, the characteristic energy
in each term should be considered as an independent parameter.  This
procedure will usually give the correct parametrization 
provided that there are no particularly strong energy dependences. 
The approximation generally works  well if the fractional 
change of the survival probability over the effective range of
integration is reasonably small.  This condition is usually satisfied
for most SNO observables.

In some cases, {\it e.g.}, when observables depend on the same
combination of oscillation parameters, the exact results for
correlations can be obtained without performing complicated
integrations over energies and over instrumental characteristics.

In summary, we  find  dependences of observables on 
oscillation parameters in terms of simple functions with 
a few fitting parameters that are determined by  
exact numerical calculations.  The fitting  parameters  represent the complicated results 
of integrations over energies and over instrumental characteristics.

\subsection{Survival probabilities, observables, and correlations}
\label{subsec:examples}

We find in this subsection the dependence of different neutrino
observables on the average survival probability. 
This is the first step in the process of deriving analytic expressions
for correlations, which was outlined in the previous subsection.
We shall see that some correlations
appear clearly even when only survival probabilities are considered.
In the Appendix, we derive expressions for the survival probabilities
and show how these expressions can be used to predict correlations
among neutrino observables.

\subsubsection{Charged current in SNO and neutrino-electron scattering in SuperKamiokande}
\label{subsubsec:ccvsescsk}

The reduced CC-event rate in the SNO detector can be written as 
\begin{equation}
{\rm [CC]} ~=~P_{\rm SNO} \cdot f_B~,  
\label{ccrate}
\end{equation}
where $P_{\rm SNO} = P_{\rm SNO}(\Delta m^2, \sin^2 2\theta, E^{th})$ is the
effective survival probability for CC events in SNO experiment. 
The $hep$ neutrinos 
do not contribute significantly to the total rate for 
an energy threshold in the likely range of $5$ to $8$ MeV and therefore $hep$ neutrinos can usually be neglected. 
We   determine the flux parameter $f_B$ from  the reduced total 
rate of events in  the SuperKamiokande neutrino-electron scattering experiment:  
\begin{equation}
{\rm [ES]}_{\rm SK} ~\equiv~  N_{\rm SK}/N_{\rm SK}^{\rm SSM},
\label{eq:defnescsk}
\end{equation}
where $N_{\rm SK}$ and $N_{\rm SK}^{\rm SSM}$ are the observed and the
SSM (BP98, see Ref.~\cite{bp98}) predicted event rates, respectively.
In the case of oscillations into active neutrinos we find
\begin{equation}
f_B ~\approx ~\frac{\rm [ES]_{\rm SK}}{P_{\rm SK} + (1 - P_{\rm SK})r}~,
\label{fB}
\end{equation}  
where $r \approx 0.16$~\cite{sirlin} is the ratio of the 
$\nu_{\mu} - e$ to the $\nu_e - e$ cross-sections,  
and $P_{\rm SK} = P_{\rm SK}(\Delta m^2, \sin^2 2\theta, E^{th})$
is the effective survival probability for the SK experiment. 

Thus, we find from Eq.~(\ref{ccrate}) and from Eq.~(\ref{fB})
\begin{equation}
{\rm [CC]}  ~\approx~
\frac{\rm [ES]_{\rm SK}}{(1 - r)(P_{\rm SK}/P_{\rm SNO}) + r/P_{\rm SNO}}~, 
\label{eq:ccandescsk}
\end{equation}
where the ratio $P_{\rm SK}/P_{\rm SNO}$ depends in general on the
oscillation parameters but is often of the order of
unity. Equation~(\ref{eq:ccandescsk}) simplifies considerably if we
make the approximation (valid for example if the SK and SNO energy
thresholds are chosen near the plausible values of $E_{\rm SK}^{th}
\sim 6.5$ MeV and $E_{\rm SNO}^{th} \sim 5$ MeV, see
Ref.~\cite{Villante}) that $P_{\rm SK} ~\approx ~P_{\rm SNO} ~=~
P$. In this special circumstance,
\begin{equation}
{\rm [CC]}  ~\approx~ \frac{\rm [ES]_{\rm SK}}{1 - r + r/P}~.                \label{R-R}
\end{equation} 
Equation~(\ref{R-R}) is generally valid for the LMA and LOW
solutions~\cite{lma}, for which the survival probability depends only
weakly on the energy in the energy range of interest.

In the case of conversion to sterile neutrinos ($r \rightarrow 0$), 
we have
\begin{equation}
{\rm [CC]_{\rm Sterile}} ~=~  {\rm [ES]}_{\rm SK}\frac{P_{\rm SNO}}{P_{\rm SK}}
~\approx ~{\rm [ES]}_{\rm SK}.  
\label{ccster}
\end{equation}
For $P_{\rm SK} ~\approx~ P_{\rm SNO}$,   
the rate [CC] is approximately equal to ${\rm [ES]}_{\rm SK}$.  

Deviations from the equality $P_{\rm SK} ~=~ P_{\rm SNO}$ can be
caused by  a strong energy dependence of the
survival probability, by differences in the energy dependences of the
neutrino cross-sections, by  difference of energy thresholds, and by
differences in  instrumental responses.

\subsubsection{Shift of the first moment of the CC spectrum}
\label{subsubsec:shiftfirst}

The fractional shift of the first moment of the recoil electron energy
spectrum [see Eq.~(\ref{eq:mom})] is easily shown (for a negligible $hep$
flux) to be proportional to the derivative of the survival
probability:
\begin{equation}
\delta T  ~\propto~ \frac{E}{P} \frac{dP}{dE}~ , 
\label{firstmom}
\end{equation}
where $P$ and $dP/dE$ are suitable spectrum averages and $E$ is a
characteristic energy.

\subsubsection{Day-Night asymmetry}
\label{subsubsec:daynight}

The day-night asymmetry  can be estimated from the expression  
\begin{equation}
A_{\rm N-D}^{\rm CC} ~=~ 2 \frac{P_{\rm N} - P_{\rm D}}{P_{\rm N} +
P_{\rm D}}~,
\label{Adn}
\end{equation}
where $P_N$ and  $P_D$  are the day and the night survival probabilities 
averaged over the year after removal of the geometrical factor $R^{-2}$. 
We showed previously in Ref.~\cite{bks2000} that   
the day-night asymmetry 
in the SNO CC-events, $A_{\rm N-D}^{\rm CC}$, and the 
asymmetry in the neutrino-electron scattering observed by
SuperKamiokande and SNO, $A_{\rm N-D}^{\rm ES}$,
are related by the approximate equation 
\begin{equation}
A_{\rm N-D}^{\rm CC} ~=~ A_{\rm N-D}^{\rm ES}
\left[1 +  \frac{r}{(1 - r)P} \right]~. 
\label{eq:A-A}
\end{equation}
 Equation~(\ref{eq:A-A}) shows that the CC day-night asymmetry that
will be measured in the SNO experiment is predicted to be larger than
the neutrino-electron scattering asymmetry measured by
SuperKamiokande. For typical values of the average survival
probability in the range $P = 0.3-0.5$, the enhancement factor in
brackets in Eq.~(\ref{eq:A-A}) is between $1.4$ and $1.6$. The
prediction given in Eq.~(\ref{eq:A-A})
can be tested also by using SNO data alone since both $A_{\rm N-D}^{\rm
CC}$ and $A_{\rm N-D}^{\rm ES}$ are measurable by SNO.
 
Combining Eqs.~(\ref{R-R}) and (\ref{eq:A-A}),  we obtain a 
relation between the  day-night asymmetry  $A_{\rm N-D}^{\rm CC}$ and 
the reduced CC rate [CC]: 
\begin{equation}
A_{\rm N-D}^{\rm CC} ~=~ \frac{A_{\rm N-D}^{\rm ES} {\rm[ES]}_{\rm
SK}}{{\rm [CC]}(1 - r)}~,  
\label{A-Rcc}     
\end{equation} 
where $A_{\rm N-D}^{\rm ES}$ depends mainly on the $\Delta m^2$ for
the LMA and LOW solution regions (see Ref.~\cite{bks2000}):  
$A_{\rm N-D}^{\rm ES} ~=~ A_{\rm N-D}^{\rm ES}(\Delta m^2)$.  This
relation holds pointwise, i.e., for a particular choice of 
$\Delta m^2$  and $\sin^2 2\theta$.  Most of the range  in the
$A_{\rm N-D}^{\rm CC}$ and [CC] plane is due to the allowed range in
$\Delta m^2$  and $\sin^2 2\theta$, which washes out the pointwise
dependence of Eq.~(\ref{A-Rcc}) because in the LMA and LOW solution
regions $A_{\rm N-D}^{\rm CC}$ depends
primarily on $\Delta m^2$ and [CC] primarily depends upon $\sin^2
2\theta$
(cf. Fig.~\ref{fig:cca5} and Fig.~\ref{fig:cca8} ).

\subsubsection{The double ratio [NC]/[CC]}
\label{subsubsec:doublenccc}

The double ratio [NC]/[CC] is equal to the inverse of the
appropriately-averaged survival probability for the active neutrino
case:
\begin{equation}
{\rm [NC]/[CC]} ~=~ \frac{1}{P}~.
\label{CC-NC}
\end{equation}
Both [NC]/[CC] and  [CC] are determined by $P$ 
[see Eq.~(\ref{R-R})]. 
Inserting Eq.~(\ref{CC-NC}) into   Eq.~(\ref{R-R}),  we obtain 
\begin{equation}
{\rm [CC]} \approx \frac{\rm [ES]_{\rm SK}}{1 - r + r {\rm
[NC]/[CC]}}~,
\label{R-NCCC}     
\end{equation}
which implies that [NC]/[CC] and [CC] are strongly correlated in our
approach (in which $f_B$ is fixed by the measured SuperKamiokande
rate). As a consequence of Eq.~(\ref{R-NCCC}), the correlation plots
are similar for [NC]/[CC] and [CC] when combined with other
observables.
A strong correlation exists also between the double ratios 
[ES]/[CC] and [NC]/[CC], both of which will be measured by SNO: 
\begin{equation}
\frac{\rm [ES]}{\rm [CC]} = 1 - r + r \frac{\rm [NC]} {\rm [CC]}~.
\label{ES-NCCC}
\end{equation}

For sterile neutrinos, 
\begin{equation}
{\rm [NC]/[CC]}_{\rm Sterile} ~=~ 
\frac{P'}{P}  ~\approx~ 1 ~,
\label{RNCCCst} 
\end{equation}
where $P'$ is the average survival probability for the NC event
sample. Since the thresholds for NC events ($2.2$ MeV) and for CC
events (expected to be greater than $5$ MeV) are different, the ratio
of the average survival probabilities, $P'/P$, is in general different
from one. However, for both NC and CC events the cross section
increases with neutrino energy and most of the events that are
observed correspond to neutrinos with relatively high energies. For
these higher energy neutrinos, the survival probability depends rather
 weakly on energy (see Fig. 1 of Ref.~\cite{bks2000}).  So, while
$P'/P$ is not identical to one it is in general quite close to one for
practical cases.

Combining Eq.~(\ref{eq:A-A}) and  Eq.~(\ref{CC-NC}), we find
\begin{equation}
{\rm [NC]/[CC]} ~=~\left( \frac{1-r}{r}\right)\Biggl[ {{A_{\rm
N-D}^{\rm CC} - A_{\rm N-D}^{\rm ES}} \over {A_{\rm N-D}^{\rm ES}}
}\Biggr]~.
\label{eq:ncccvsa}
\end{equation}
Equation~(\ref{eq:ncccvsa}) is an example of a correlation between
three observables. The equality given in Eq.~(\ref{eq:ncccvsa}) does
not depend explicitly on the oscillation parameters and holds
approximately for all three MSW active neutrino solutions.  The
principal inaccuracy introduced in the derivation of
Eq.~(\ref{eq:ncccvsa}) is caused by the fact that the average survival
probability, $P$, that appears in Eq.~(\ref{eq:A-A}) is not exactly
equal to the average survival probability that appears in
Eq.~(\ref{CC-NC}).

If we neglect the dependence of the measured quantities upon energy
threshold and upon the energy dependence of the rates, then the
neutrino observables depend only on two parameters, $\Delta m^2$ and
$\sin^2 2\theta$ . Therefore any three observables $X, Y, Z$ must be
correlated, except for special cases in which one or more of the
observables do not depend on the oscillation parameters. Indeed,
expressing the oscillation parameters in terms of the two observables,
say $X$ and $Y$, we can get the relation $Z = Z(\Delta m^2, \sin^2
2\theta) = Z(X, Y)$.  The experimental study of the validity of
Eq.~(\ref{eq:ncccvsa}), and other similar ``triple" relations, will
provide important tests of the consistency of the oscillation
solutions and the experimental results. Deviations from the ``triple"
relations that could not be explained by expected energy dependences
of the experimental quantities, or by differences in the average
values of $P$ for the various measurables, 
would indicate either the participation of more than two
neutrinos in solar neutrino oscillations or a lack of consistency of
the experimental results.

\subsubsection{What's next?}
\label{subsubsec:whatnext}
 
The functional dependence of the survival probability on the
oscillation parameters depends on which particular solution of the
solar neutrino problems is chosen. In the Appendix, we give the
function dependences for different currently-favored oscillation
scenarios.  Using the expressions for $P$ given in the Appendix and
the relations presented in Eq.~(\ref{R-R}), Eq.~(\ref{firstmom}),
Eq.~(\ref{Adn}), and Eq.~(\ref{CC-NC}), we derive the dependences of
the SNO observables on the neutrino oscillation parameters.

In the next three sections, we present maps of neutrino oscillation
solution regions onto planes constructed from different pairs of SNO
observables.  We discuss results for the following currently-favored
two-neutrino solutions which explain all of the available solar
neutrino data: Large Mixing Angle (LMA) MSW solution, Small Mixing
Angle (SMA) MSW solution, Low $\Delta m^2$ (LOW) solution, and MSW
Sterile solution based on small mixing angle MSW conversion to sterile
neutrinos.  There are several disconnected regions (``islands") of the
Vacuum Oscillation solutions. We will divide them into two groups:
vacuum oscillation solutions with small $\Delta m^2$ ($\Delta m^2 <
10^{-10}$ eV$^2$), VAC$_{\rm S}$, and several islands with large
$\Delta m^2$ ($\Delta m^2 > 10^{-10}$ eV$^2$) solutions, VAC$_{\rm
L}$.  For VAC$_{\rm S}$, there are currently four allowed islands in
the $\Delta m^2$ and $\sin^2 2\theta$ plane. The VAC$_{\rm S}$
solutions correspond to four almost fixed values of $\Delta m^2$ and
varying $\sin^2 2\theta$.

\section{CC-rate versus Day-Night Asymmetry}
\label{sec:r-a}

\begin{figure}[!t]
\centerline{\psfig{figure=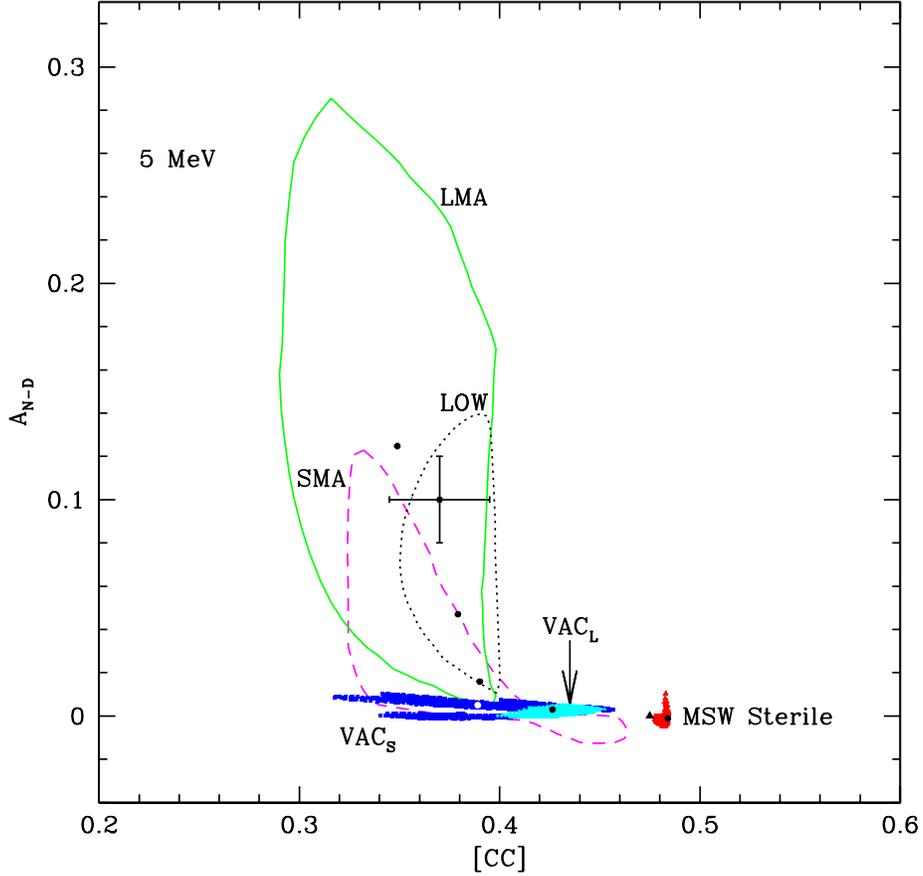,width=5.0in,angle=0}}
\tightenlines
\caption[]{\small The allowed regions for the day-night asymmetry in
the charged current event rate,
$A_{\rm N-D}$, versus the reduced charged current rate, [CC], for an 
electron energy threshold 5 MeV.  The figure shows the
currently-allowed regions predicted by two-neutrino
solutions~\cite{snoshow} that describe all the available solar
neutrino data: LMA (encircled by a solid line), SMA (dashed line), LOW
(dotted line), VAC$_{\rm S}$ (black points), and VAC$_{\rm L}$ (grey
points).  The best-fit points for each solution are indicated by a
small black circles within the allowed region.  The prediction for the
no-oscillations case is indicated by a triangle.  The cross near the
best-fit point of the LMA solution is a simulated measurement with
estimated $1 \sigma$ error bars.
\label{fig:cca5}}
\end{figure}

\begin{figure}[!ht]
\centerline{\psfig{figure=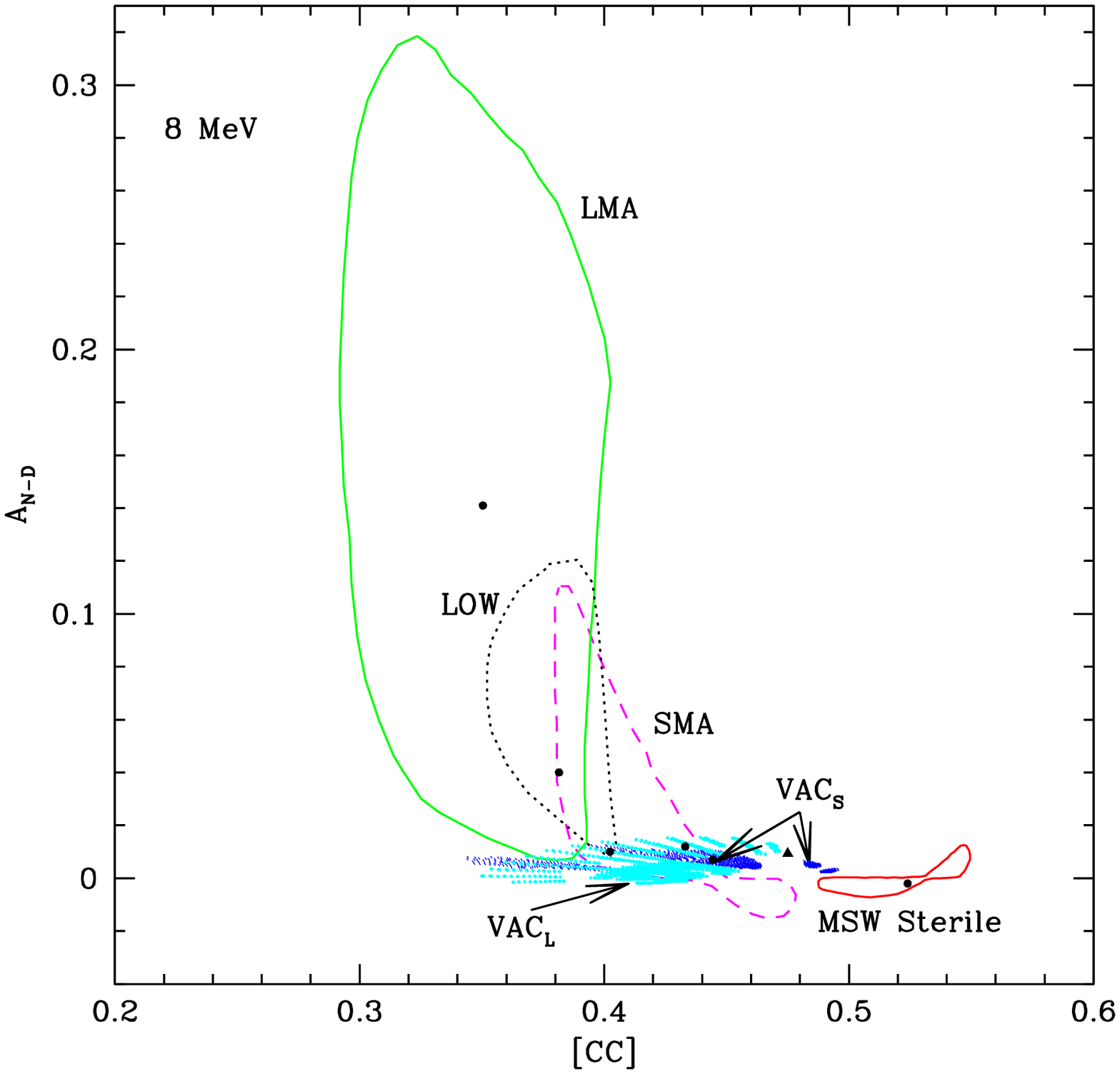,width=5.0in,angle=0}}
\tightenlines
\caption[]{\small The allowed regions for the day-night asymmetry,
$A_{\rm N-D}$, versus the reduced charged current rate, [CC], for an
electron energy threshold $8$ MeV.  The meaning of the symbols is the
same as for Fig.~\ref{fig:cca5}, except that the regions now refer to a
recoil electron energy threshold of 8 MeV.
\label{fig:cca8}}
\end{figure}

Figures~\ref{fig:cca5} and \ref{fig:cca8} show maps of the
currently-allowed regions in the $\Delta m^2 - \sin^2 2\theta$ plane
onto the ${\rm [CC]} - A_{\rm N-D}$ plane for the electron energy
thresholds of 5 MeV and 8 MeV, respectively. In Fig.~\ref{fig:cca5},
we show a simulated data point near the best-fit value for the LMA
solution.  The estimated $1\sigma$ error bar for the CC measurement is
taken from Table~II of Ref.~\cite{bks2000}.  For $A_{\rm N-D}$, we
assume for purposes of illustration a $\pm 0.03$ uncertainty in the
absolute value as a $1\sigma$
error, which is comparable with the accuracy that has been achieved
after three years with the SuperKamiokande detector~\cite{superk}.

\subsection{Discriminating among solutions}
\label{subsec:ccvsadiscriminating}

There are four regions in Fig.~\ref{fig:cca5} and Fig.~\ref{fig:cca8}
for which (after taking account of the likely measurement uncertainties
and the overlap of the predicted values of the observables)
important scientific inferences will be possible if the measured
values of [CC] and $A_{\rm N-D}$ fall within the designated areas. 1)
If measurements show that $A_{\rm N-D} > 0.2$, then that will be a
strong indication in favor of the LMA solution. 2) If the measurements
only show that $A_{\rm N-D} > 0.1$, that by itself will be sufficient
to disfavor the vacuum and MSW Sterile solutions.  3) If instead
$A_{D-N} \sim 0$ and [CC] is consistent with $0.48$, then that would
strongly favor the MSW Sterile solution. 4) If the measured values lie
within the `zone of avoidance' of $A_{\rm N-D} > 0.02$ and ${\rm [CC]}
> 0.4$, then none of the currently acceptable oscillation solutions
will be favored.

If $A_{\rm N-D} > 0.1$, then the inferences from cases 1) and 2) above
can be tested by measuring $\delta T$. Figure~\ref{fig:ccm5} and
Fig.~\ref{fig:ccm8}, as well as Fig.~\ref{fig:ma5} and
Fig.~\ref{fig:ma8},  show that all the currently allowed oscillation
solutions predict $\delta T < 0.01$ if $A_{\rm N-D} > 0.15$.

If the day-night asymmetry lies in the broad range $0 < A_{\rm N-D}
<0.15$, it will be difficult to disentangle the LMA, LOW and SMA
solutions.  These three MSW solutions show a large overlap in their
predictions for the [CC]-$A_{\rm N-D}$ plane (see Fig.~\ref{fig:cca5}
and Fig.~\ref{fig:cca8} and especially Table~IX of
Ref.~\cite{bks2000}).  If the observed asymmetry is not too small,
e.g., if $A_{\rm N-D} > 0.1$, then the SMA solution can be identified
by the zenith angle dependence of the rate during the night.
According to the SMA scenario, the rate should be strongly enhanced in
the deepest night bin (for the core-crossing neutrino trajectories)
\cite{core}.  In contrast, the LMA and LOW solutions predict rather
flat zenith angle distributions. The LOW and the LMA solutions may be
distinguishable through the observed dependence of the day-night
asymmetry on the energy threshold.  The asymmetry increases with
threshold for the LMA solution and decreases with threshold for the
LOW solution.  For the LMA solution, the maximal possible asymmetry
becomes as large as 0.32 for $E^{th} = 8$ MeV instead of 0.28 for $E =
5$ MeV (see Fig.~\ref{fig:cca5} and Fig.~\ref{fig:cca8}).  For the LOW
solution, the dependence upon the threshold energy is just the
opposite; the predicted asymmetry decreases with increasing threshold
energy. In particular, the LOW solution predicts that the maximal
asymmetry decreases from $0.135$ to $0.11$ as the threshold energy
increases from $5$ MeV to $8$ MeV. The LOW solution may also be identified
later by  strong Day-Night variations of the beryllium line
in BOREXINO experiment \cite{borexino}.

The charged-current event ratio is in some ways the simplest
experimental quantity to measure with the Sudbury Neutrino
Observatory. However, the most remarkable aspect of the above analysis
of Fig.~\ref{fig:cca5} and Fig.~\ref{fig:cca8} is that the potentially
important inferences are almost entirely independent of the measured
charged-current rate. This is because the estimated $1\sigma$
uncertainty in the value of [CC] is about $6.7$\% (see
Ref.~\cite{bks2000}) and is dominated by the theoretical uncertainty
in the charged-current neutrino-absorption cross section.  
Unless a major improvement is made in the
accuracy of the theoretical cross section calculation, the potential
diagnostic value of the charged-current measurement will be severely
compromised by the large uncertainty in the neutrino absorption cross
section.

\subsection{Correlation phenomenology}
\label{subsec:correlationphenom}

For the LMA solution, there is no significant correlation shown in
Fig.~\ref{fig:cca5} and Fig.~\ref{fig:cca8}; the correlation
parameter, cf. Eq.~(\ref{eq:kappa}), $\kappa_{A-CC}({\rm LMA}) \sim 1$.  
For the largest area of the plane of oscillation parameters, the
charged-current rate depends mainly on $\sin^2 2\theta$ [see
Eq.~(\ref{P-lma})], whereas $A_{\rm N-D}$ depends strongly on $\Delta
m^2$ [see Eq.~(\ref{A-lma})].  There is a tendency for small values of
[CC] to correspond to large values of $A_{\rm N-D}$, since [according
to Eq.~(\ref{A-Rcc})] for fixed $\Delta m^2$ the asymmetry is
inversely proportional to [CC].

Also for the LOW solution, no significant correlation appears.  The
area occupied in the [CC]-$A_{\rm N-D}$ plane by the LOW solution is
substantially smaller than for the LMA solution, which reflects the
smaller allowed region of the LOW solution in the $\Delta m^2 - \sin^2
2 \theta$ plane.

The SMA solution has the form of two beautiful, asymmetric petals
connected at the point of zero asymmetry.  This form can be understood
from the expression for the asymmetry, Eq.~(\ref{A-sma}).  The zero
asymmetry contour is determined by the condition $P = 1/2$.
Therefore, the contours of $A_{\rm N-D} = 0$ and of ${\rm [CC]} =
0.41$, both of which correspond to $P = 1/2$, coincide. The contours
are defined by the relation
\begin{equation}
\xi \equiv \Delta m^2 \cdot \sin^2 2\theta  =   {\rm constant}~ . 
\label{eq:xiparam}
\end{equation}
The correlation between $A_{\rm N-D}$ and 
[CC] appears in the region of small asymmetries and of large [CC].  
The rate [CC] decreases with increase of $A_{\rm N-D}$  
(Here $\kappa_{A-CC}({\rm SMA}) \ll 1$ )
\footnote{A similar plot for correlation of the slope parameter and the
asymmetry have been given in Ref.~\cite{AS}.}.

For vacuum solutions, there is a correlation between $A_{\rm N-D}$ and
[CC] that is difficult to see on the scale of Fig.~\ref{fig:cca5} and
Fig.~\ref{fig:cca8}, because the day-night asymmetry is small.  The
residual asymmetry, which is calculated after first removing the
$R^{-2}$ dependence of the total flux, is not zero, but $A_{\rm N-D} <
2\%$ is predicted for all the currently-favored vacuum oscillation
solutions~\cite{bks2000}. By inspecting the figures carefully, one can
see that the asymmetry $A_{\rm N-D}$ increases with decreasing [CC].

The day-night effect for vacuum oscillations is determined by
geometrical factors.  In the northern hemisphere, the nights are
longer in the winter when the earth is closer to the sun. The
earth-sun distance affects the vacuum oscillation probability and
therefore the CC event rate.
Combining Eq.~(\ref{R-R}), Eq.~(\ref{P-vac}), and Eq.~(\ref{A-vac}),
we find
\begin{equation}
A_{\rm N-D} = \frac{1}{r}\left(\frac{\rm [ES]_{\rm SK}}{\rm [CC]} -
1\right)\times f(\Delta m^2)~,
\label{A-RCC}
\end{equation}
where 
\begin{equation}
f(\Delta m^2) \approx 2 (\Delta m^2/m_V^2) \cot (\Delta m^2/m_V^2)
\label{fdef}
\end{equation}
is a function of $\Delta m^2$ only.  Since within a given
currently-allowed ``island" in the $\Delta m^2 - \sin^2 2\theta$ plane
, the variation of $\Delta m^2$ is small, we can consider $f(\Delta
m^2) \approx {\rm constant}$. Equation~(\ref{A-RCC}) explains the
correlation between $A_{\rm N-D}$ and [CC] that exists in
Fig.~\ref{fig:cca5} and Fig.~\ref{fig:cca8}.

\section{Charged-Current Rate versus Shift of First Moment}
\label{sec:r-mom}

Figures~\ref{fig:ccm5} and \ref{fig:ccm8} show, for electron
energy thresholds of $5$ MeV and $8$ MeV, maps onto the ${\rm [CC]} -
\delta T$ plane of the currently-allowed regions in the $\Delta m^2 -
\sin^2 2\theta$ plane.  For illustrative purposes,
Fig.~\ref{fig:ccm5} shows a simulated experimental point near  the
current best-fit predicted point for the LMA solution. The error bars
are estimated $1\sigma$ uncertainties from Table~II of
Ref.~\cite{bks2000}, with a $1.3$\% fractional uncertainty in the
first moment and a $6.7$\% for the charged-current rate.

\subsection{Discriminating among solutions}
\label{subsec:ccvsdeltat}

If the measured values of [CC] and $\delta T$ fall close to the
current best-fit value of the LMA solution, then many of the
currently-favored solutions will still be allowed if a $3\sigma$ level
of disagreement is permitted. The difficulty in uniquely identifying
solutions is primarily caused by the estimated $3\sigma$ uncertainties
being comparable in many cases to the size of the predicted effects.

\begin{figure}[!t]
\centerline{\psfig{figure=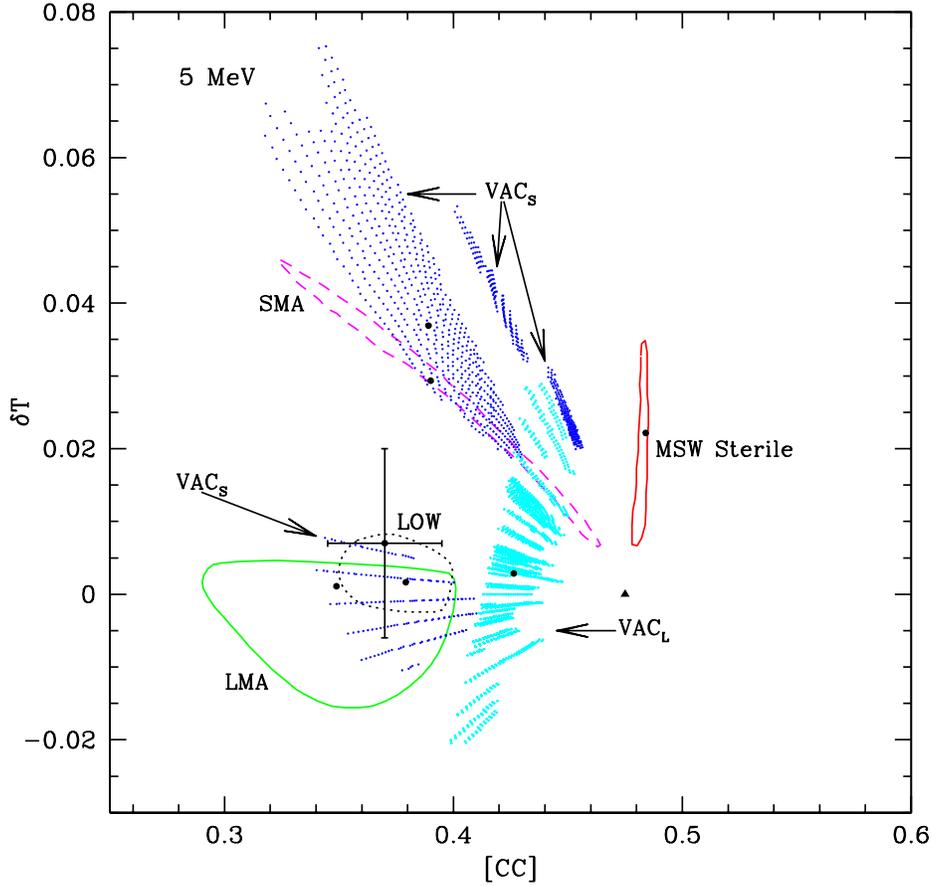,width=5.0in,angle=0}}
\tightenlines
\caption[]{\small The allowed regions for the shift of the first
moment, $\delta T$, versus the reduced charged current rate, [CC], for
a recoil electron energy threshold of 5 MeV.  The meaning of the
symbols is the same as in Fig.~\ref{fig:cca5}, except that the regions
now refer to $\delta T$ and [CC]. The nearly horizontal segments that
overlap with the LOW and the LMA solution regions correspond to
VAC$_{\rm S}$ solutions with the largest values of $\Delta m^2$.
\label{fig:ccm5}}
\end{figure}

There are some regions of the two-dimensional parameter space, ${\rm
[CC]} - \delta T$, that are relatively discriminatory. For example, in the
region in which $\delta T > 0.04$ and $0.3 < {\rm [CC]} <0.4$, only
the VAC$_S$ and the SMA solutions are represented. The most extreme
values of the [CC] parameter, e.g., [CC] $> 0.5$ or [CC] $< 0.3$,
would indicate, respectively, the MSW Sterile solution or the LMA
solution.  If either of these cases is suggested by the [CC]
measurement, then a comparison of the predicted and measured $\delta
T$ (Fig.~\ref{fig:ccm5}) and $A_{\rm N-D}$ (Fig.~\ref{fig:cca5}) will
be useful checks of the validity of the identification of the
solution.

Unique inferences will be possible
(see Fig.~\ref{fig:ccm5}) for extreme  VAC$_S$ solutions with a
fractional shift $\delta T > 4.5$\%  and for the  MSW
Sterile solution with [CC] greater than 0.48. The extreme VAC$_S$
solution predicts a very small value for the day-night asymmetry,
$A_{\rm D-N} < 0.01$ (see Fig.~\ref{fig:ma5}).

\begin{figure}[!ht]
\centerline{\psfig{figure=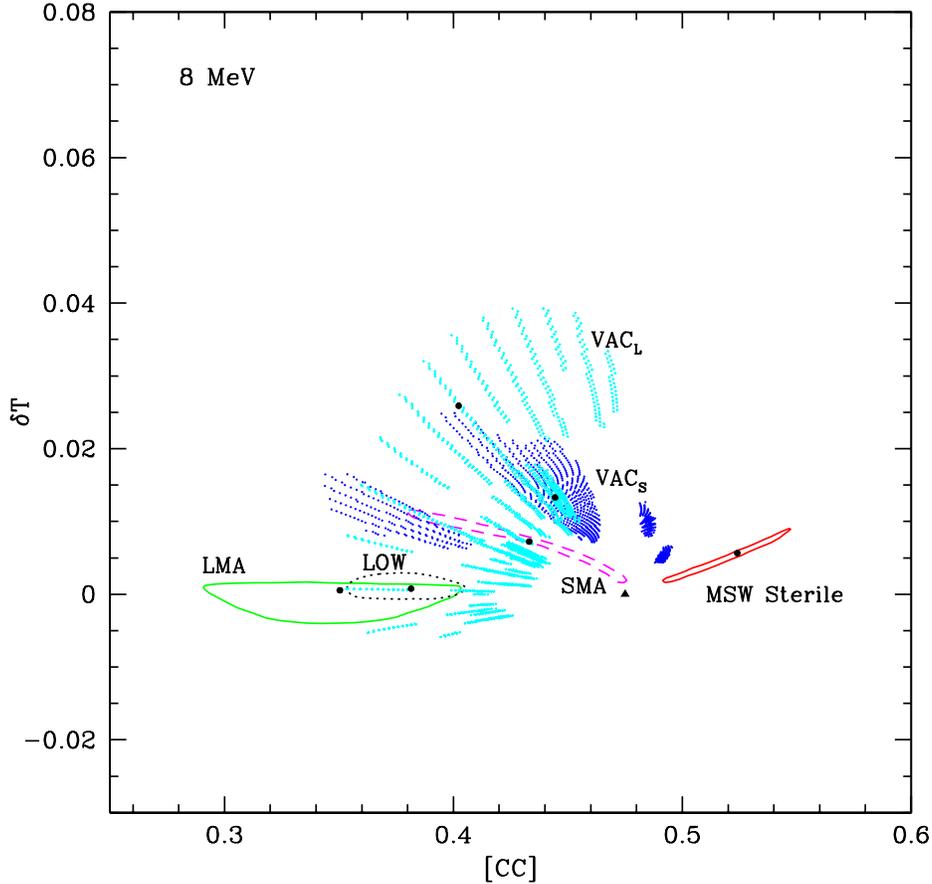,width=5.0in,angle=0}}
\tightenlines
\caption[]{\small The allowed regions for the shift of the first
moment, $\delta T$, versus the reduced charged current rate, [CC], for
a recoil electron energy threshold of $8$ MeV.  The meaning of the
symbols is the same as in Fig.~\ref{fig:cca5},  except that the allowed
regions refer to $\delta T$ and [CC] and the threshold for the recoil
electron energy is $8$ MeV.
\label{fig:ccm8}}
\end{figure}

Two zones of avoidance appear in Fig.~\ref{fig:ccm5} for large
[CC]. None of the currently favored oscillation solutions predict
$\delta T < 0.01$ or $\delta T > 0.04$ for [CC] $> 0.45$. For an
electron recoil energy threshold of $8$ MeV, we find that 
$\delta T < 0.04$ for all
the currently favored oscillation solutions (see Fig.~\ref{fig:ccm8}).
For smaller [CC], between $0.3$ and $0.4$, 
there is also a zone of avoidance for $\delta T > 0.01$ and less than
the values predicted by the SMA solution.

\subsection{Correlation phenomenology}
\label{subsec:ccvsdeltatphenom}

For the SMA solution, both [CC] and $\delta T$ are determined by a
unique combination of neutrino variables, $\xi$, defined by
Eq.~(\ref{eq:xiparam}), so that (up to small earth matter effect
corrections) the two measurables are strongly correlated. As $\delta
T$ increases, the charged-current rate [CC] decreases
(cf. Fig.~\ref{fig:ccm5}). Using Eq.~(\ref{eq:ccandescsk}) (with
$P_{\rm SNO} \sim P_{\rm SK}$) and Eq.~(\ref{r-sma}) for the rate and
Eq.~(\ref{mompar}) for the shift of the first moment, we find for an
electron threshold energy of $5$ MeV
\begin{equation}
{\rm [CC]} \approx \frac{\rm [ES]_{\rm SK}}{1 - r + r \cdot e^{B \delta T} }~.
\label{CCmomcorr}
\end{equation}
The numerical coefficient $B$ in the exponent that occurs in
Eq.~(\ref{CCmomcorr}) is determined by results of exact numerical
calculations, $B = 29.3$.  For an electron energy threshold of $8$
MeV, one should use the general formula given in
Eq.~(\ref{eq:ccandescsk}) without making the approximation that
$P_{\rm SNO} \sim P_{\rm SK}$.

For the VAC$_{\rm S}$ solution, a strong correlation exists.  Using
Eq.~(\ref{R-R}) and Eqs.~(\ref{mom-vac}) and (\ref{P-vac}), we find
\begin{equation}
\delta T = \frac{1}{r}\left(\frac{\rm [ES]_{\rm SK}}{\rm [CC]} -
1\right)\times f(\Delta m^2)~,
\label{T-RCC}
\end{equation} 
where the function $f(\Delta m^2)$, has been defined by
Eq.~(\ref{fdef}).  The relation given in Eq.~(\ref{T-RCC}) holds for
each allowed island in neutrino parameter space, and, in the
approximation of constant $\Delta m^2$, there is a strong correlation
of the rate and the shift of the first moment for each of the three
islands with low $\Delta m^2$.  The first moment shift, $\delta T$,
increases as [CC] decreases.  For the allowed island with the largest
value of $\Delta m^2$ (which overlaps for a $5$ MeV threshold with the
LMA and LOW solutions), the shift of the first moment is close to zero
and the correlations are weak.

For the MSW Sterile neutrino solution, the rate [CC] is strongly
restricted by the measured value of ${\rm [ES]}_{\rm SK}$, whereas
$\delta T$ varies over a significant range.

The LMA solution does not predict a strong correlation, as can easily
be seen from Eqs.~(\ref{P-lma}) and (\ref{P-mom}).  The rate [CC]
depends strongly on $\sin^2 2\theta$, whereas $\delta T$ depends
strongly on $\Delta m^2$.  The situation is similar for the LOW
solution.

At a higher threshold, $8$ MeV (see Fig.~\ref{fig:ccm8}), the shift of
the first moment becomes smaller for all solutions. In particular, a
significant part of the LMA region with negative $\delta T$
disappears. At the same time, for the VAC$_{\rm L}$ region the
best-fit point shifts to larger $\delta T$. In contrast, the spread of
the [CC] rates increases especially for the SMA Sterile solution. For
$E^{th} = 5$ MeV, $P_{SNO} \approx P_{SK}$ and, as a consequence, [CC]
is uniquely fixed by $R_{SK}$ [Eq.~(\ref{ccster})].  For $E^{th} = 8$
MeV, $P_{SNO}/ P_{SK}$ differs from one and depends on oscillation
parameters, which leads to the larger spread in [CC].  Also for
$E^{th} = 8$ MeV, the VAC$_{\rm S}$ region with the largest $\Delta
m^2$ no longer overlaps with the LMA and the LOW solution regions.

\section{Shift of First Moment versus Day-Night Asymmetry}
\label{sec:a-mom}

\begin{figure}[!t]
\centerline{\psfig{figure=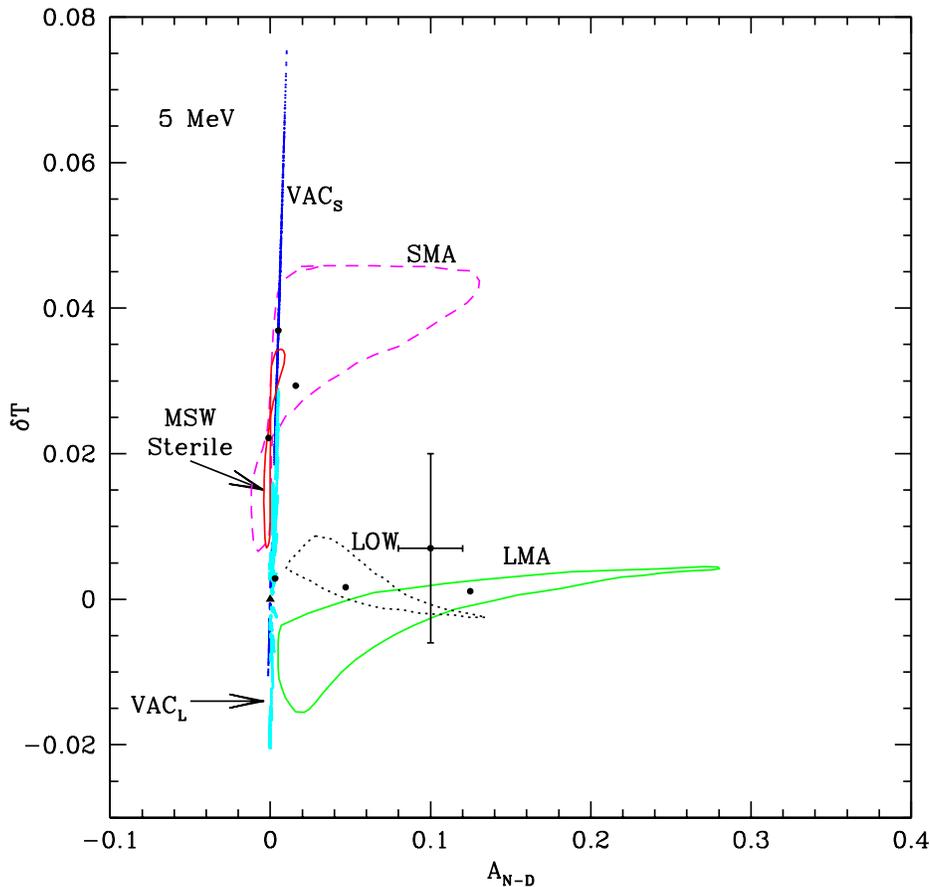,width=5.0in,angle=0}}
\tightenlines
\caption[]{\small The allowed regions for the shift of the first
moment, $\delta T$, versus the day-night asymmetry, $A_{\rm N-D}$, for
an electron recoil electron energy threshold of $5$ MeV. The meaning
of the symbols is the same as in Fig.~\ref{fig:cca5}, except that the
allowed regions now refer to $\delta T$ and $A_{\rm N-D}$.   
\label{fig:ma5}}
\end{figure}

\begin{figure}[!ht]
\centerline{\psfig{figure=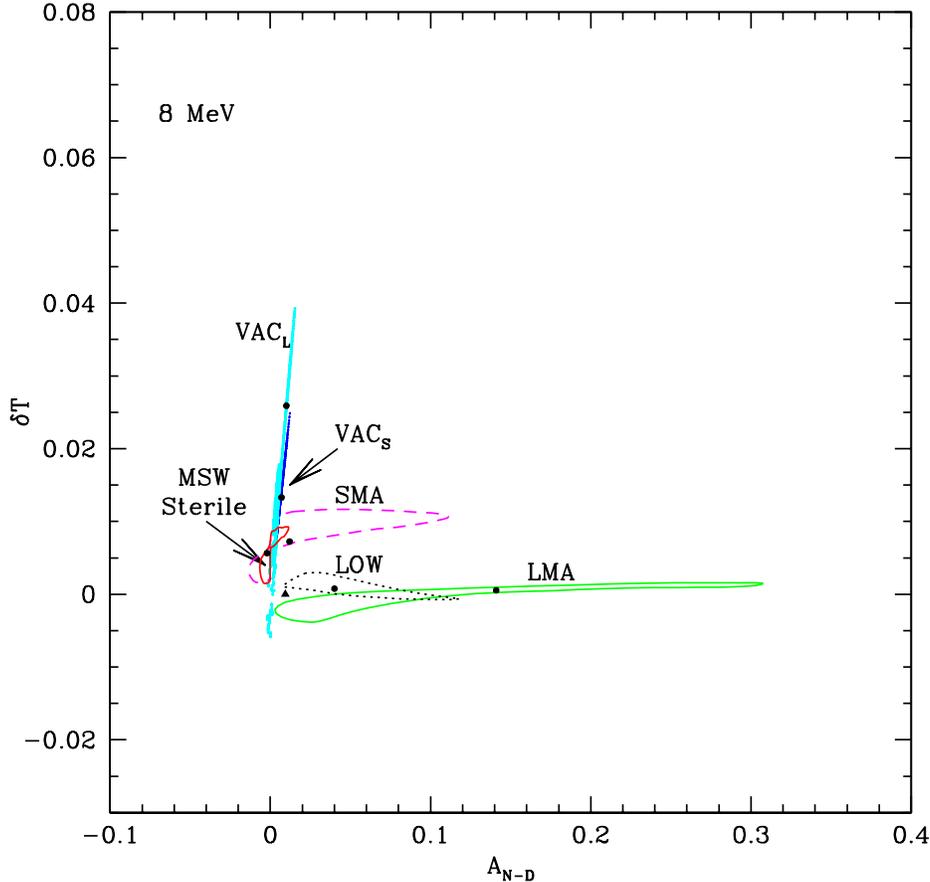,width=5.0in,angle=0}}
\tightenlines
\caption[]{\small 
The allowed regions for the shift of the first
moment, $\delta T$, versus the day-night asymmetry, $A_{\rm N-D}$, for
an electron recoil electron energy threshold of $8$ MeV. The meaning
of the symbols is the same as in Fig.~\ref{fig:cca5}, except that the
allowed regions now refer to $\delta T$ and $A_{\rm N-D}$ and the 
threshold for the recoil electron energy is $8$ MeV. 
\label{fig:ma8}}
\end{figure}

Figures~\ref{fig:ma5} and \ref{fig:ma8} show maps of the
currently-allowed regions in the $\Delta m^2 - \sin^2 2\theta$ plane
onto the $\delta T - A_{\rm N-D}$ plane for the electron energy
thresholds of 5 MeV and 8 MeV, respectively. In Fig.~\ref{fig:ma5},
the estimated $1.3$\% ($1\sigma$) error bar for the $\delta T$
measurement is taken from Table~II of Ref.~\cite{bks2000}.  For
$A_{\rm N-D}$, we assume for purposes of illustration a $\pm 0.03$
uncertainty in the absolute value as a $1\sigma$ error, 
which is comparable with the
accuracy that has been achieved after three years with the
SuperKamiokande detector~\cite{superk}.

\subsection{Discriminating among solutions}
\label{subsec:avsdeltat}

The only truly unique regions in the $A_{\rm N-D} - \delta T$ plane
are the very large values of $A_{\rm N-D} > 0.2$, which would favor
LMA, and the very large values of $\delta T > 0.06$ ($5$ MeV
threshold), which would favor VAC$_S$. In both cases, the oscillation
solution implies that the other measured parameter should be small,
i.e., $\delta T$ should be small (according to LMA) if $A_{\rm N-D} $
is near its maximal value and $A_{\rm N-D} $ should be small if
$\delta T$ is near its maximal value. 

Both the LMA and the LOW solutions predict that the shift of the first
moment is small for all allowed values  of the day-night
asymmetry, i.e., $-0.02 < \delta T <0.015$.  And, of course, the
day-night asymmetry is predicted to be small for all the vacuum
solutions and the MSW Sterile solution, i.e., $|A_{\rm N-D}| < 0.02$
for all allowed values of $\delta T$. The imposition of these cross
checks can be used to test the validity of currently-allowed
oscillation solutions.

Taking into account the estimated uncertainties in the measurements,
the most populated region in the $A_{\rm N-D} - \delta T$ plane
contains multiple currently-allowed solutions.

It is easy to find zones of avoidance in Fig.~\ref{fig:ma5} and
Fig.~\ref{fig:ma8}. For a $5$ MeV electron recoil energy threshold,
there are no predicted solutions with $A_{\rm N-D} > 0.02$ and $\delta
T > 0.045$ nor are there any predicted solutions with $A_{\rm N-D} >
0.1$ and $\delta T < 0$. For an $8$ MeV energy threshold, there are
no predicted solutions with $A_{\rm N-D} > 0.02$ and $\delta T >
0.01$.

\subsection{Correlation phenomenology}
\label{subsec:correlavsdeltat}

In the small $\Delta m^2$ limit (large day-night asymmetry), 
both the LMA and the LOW solutions predict an approximately linear
relation between the day-night asymmetry and the fractional shift in
the first moment:
\begin{equation}
\delta T ~=~ k A_{\rm N-D}~. 
\label{A-mom2}
\end{equation}
For the LMA solution, $k_{\rm LMA} = 0.014$ and $k_{\rm LOW} = -0.03$
for a $5$ MeV threshold.  These results can be obtained from
Eq.~(\ref{P-mom}) and Eq.~(\ref{A-lma}) for the LMA solution and from
Eq.~(\ref{mom-low}) and Eq.~(\ref{A-low}) for the LOW solution. This
weak correlation exists because in the region of small $\Delta m^2$ of
the LMA solution and large $\Delta m^2$ of the LOW solution both the
day-night asymmetry and the shift of the first moment are induced by
the earth matter effect.

The allowed regions for the SMA and the MSW Sterile solutions both have the
form of two petals connected at the point $A_{\rm N-D} = 0$.

For vacuum solutions, with the accuracy that is apparent in
Fig.~\ref{fig:ma5} and Fig.~\ref{fig:ma8}, the shift of the first
moment does not seem to depend significantly upon the day-night
asymmetry. However, there is a  linear correlation,
\begin{equation}
A_{\rm D-N}  ~\propto~ \delta T,
\label{eq:A-mom1}
\end{equation}
where the coefficient of proportionality is sufficiently small that
the variation of $A_{\rm D-N}$ about zero is not easily visible on the
scale shown in the figures.  
The correlation arises because for vacuum oscillations $A_{\rm D-N}
~\propto~ R dP/dR$ and $\delta T ~\propto~ E dP/dE$, where the
survival probability, $P$, depends upon the ratio of distance, $R$, to
energy, $E$, i.e., $P = P(R/E)$ [see discussion
following Eq.~(\ref{mom-vac}) and
Eq.~(\ref{A-vac})].
It will be very difficult to test
experimentally whether the relation given in Eq.~(\ref{eq:A-mom1}) is
present.

\section{[NC]/[CC] versus Day-Night Asymmetry}
\label{sec:nc}

\begin{figure}[!t]
\centerline{\psfig{figure=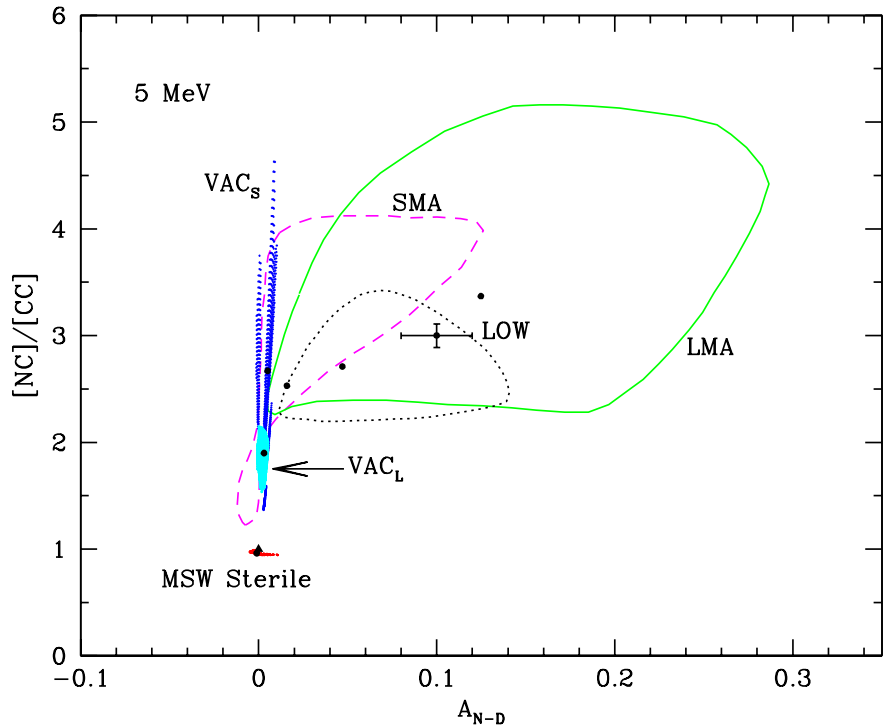,width=5.0in,angle=0}}
\tightenlines
\caption[]{\small The allowed regions for the [NC]/[CC] double ratio
versus the day-night asymmetry, $A_{\rm N-D}$, for a recoil electron
energy threshold of $5$ MeV.  The meaning of the symbols is the same
as in Fig.~\ref{fig:cca5}, except that the allowed regions now refer to
[NC]/[CC] and $A_{\rm N-D}$.

\label{fig:nca5}}
\end{figure}

\begin{figure}[!ht]
\centerline{\psfig{figure=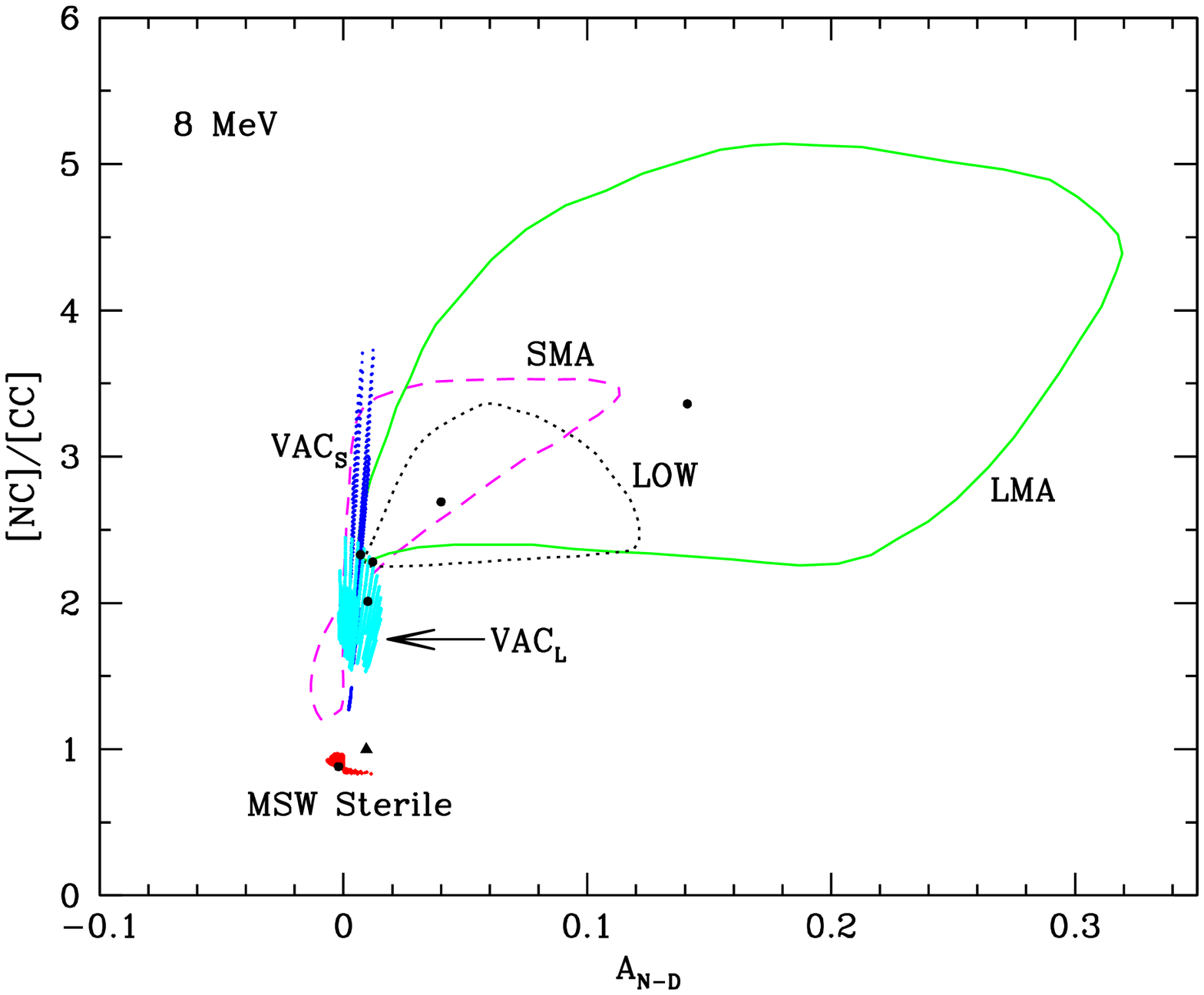,width=5.0in,angle=0}}
\tightenlines
\caption[]{\small The allowed regions for the [NC]/[CC] double ratio
versus the day-night asymmetry, $A_{\rm N-D}$, for a recoil electron
energy threshold of $8$ MeV.  The meaning of the symbols is the same
as in Fig.~\ref{fig:cca5}, except that the allowed regions now refer to
[NC]/[CC] and $A_{\rm N-D}$ and a recoil energy threshold of $8$ MeV.
\label{fig:nca8}}
\end{figure}

Figures~\ref{fig:nca5} and \ref{fig:nca8} show maps of the
currently-allowed regions in the $\Delta m^2 - \sin^2 2\theta$ plane
onto the ${\rm [NC]/[CC]} - A_{\rm N-D}$ plane for the electron energy
thresholds of 5 MeV and 8 MeV, respectively. In Fig.~\ref{fig:nca5},
we show a simulated data point near the best-fit value for the LMA
solution.  The estimated $1\sigma$ error bar for the [NC]/[CC]
measurement is $3.6$\% after one year (Table~II of
Ref.~\cite{bks2000}) and is dominated by the statistical error in the
determination of the neutral-current rate.  For $A_{\rm N-D}$, we
assume for purposes of illustration a $\pm 0.03$ uncertainty in the
absolute value as a $1\sigma$
error, comparable to  the accuracy that has been achieved
after three years with the SuperKamiokande detector~\cite{superk}.
  
The precision with which both the [NC]/[CC] ratio and the day-night
asymmetry are measured will improve with time as more events are
detected. 

Figure~\ref{fig:nca5} and Fig.~\ref{fig:nca8} are similar to
Fig.~\ref{fig:cca5} and Fig.~\ref{fig:cca8}, because of the relation
given in Eq.~(\ref{R-NCCC}) between[NC]/[CC] and [CC]. The form of the
allowed regions, the shape of the zones of avoidance, and the degree
of overlap between different solutions are all quite similar in both
sets of figures.  But, because the double ratio [NC]/[CC] can
potentially be measured with much better accuracy than [CC] alone, the
correlations of [NC]/[CC] with $A_{\rm N-D}$ and other observables
will have much stronger discriminatory power.

\subsection{Discriminating among solutions}
\label{subsec:ncccvsa}

There are regions in the [NC]/[CC] $ - A_{\rm N-D}$ plane in which
only one oscillation solution is predicted to exist and,  on the other
hand, regions
in which there is significant overlap between different oscillation
solutions. There are also significant regions, easily visible to the
eye in Fig.~\ref{fig:nca5} and Fig.~\ref{fig:nca8}, in which no
oscillation solutions are predicted to lie.

We begin with a discussion of the regions where the identification of
the oscillation solution may be unique and then discuss the ambiguous
regions and the excluded regions.

If $A_{\rm N-D}$ is observed to be greater than $0.2$ and [NC]/[CC] is
larger than $2.5$, then the LMA solution will be uniquely singled
out. The measurement of the first moment of the electron recoil energy
distribution will provide a check on this inference since the LMA
solution implies $|\delta T| < 0.01$ (see Fig.~\ref{fig:ma5} and
Fig.~\ref{fig:ma8}), i.e., a very small distortion of the
charged-current energy spectrum. Moreover, the reduced [CC] rate 
should be consistent with a value in the range $0.3$ to $0.4$
(see Fig.~\ref{fig:cca5} and Fig.~\ref{fig:cca8}).

If [NC]/[CC] is measured to be larger than $4.5$, then the only
candidate solutions will be LMA and VAC$_{\rm S}$. The two
possibilities can be distinguished since (see Fig.~\ref{fig:nca5} and
Fig.~\ref{fig:nca8}) LMA predicts a significant day-night asymmetry,
$A_{\rm N-D} > 0.06$, for large [NC]/[CC] and VAC$_{\rm S}$ predicts a
very small asymmetry, $|A_{\rm N-D}| <  0.01$.

If, on the other hand, [NC]/[CC] is found to be smaller than $2.0$,
then the LMA and LOW solutions will be eliminated. Values of [NC]/[CC]
in the range $2.0$ to $1.2$ can be obtained with the SMA and vacuum
solutions, but a value of [NC]/[CC] consistent with unity (and a small
measurement error) would uniquely favor the MSW Sterile solution.  In
all cases of [NC]/[CC] less than $2.0$, the predicted day-night
asymmetry is very small, $|A_{\rm N-D}| < 0.02$ (see
Fig.~\ref{fig:nca5} and Fig.~\ref{fig:nca8}).

The most ambiguous region will be $2 < {\rm [NC]/[CC]} < 4$ and a
small ($ < 0.02$) day-night asymmetry. Figure~\ref{fig:nca5} shows that multiple
oscillation solutions can give rise to observables in this
region. Moderate values of $A_{\rm N-D}$ (e.g., $0.02$ to $0.12$) and
moderate values of [NC]/[CC] (e.g., $2.5$ to $4.0$) will also be
ambiguous since all three of the MSW active solutions, LMA, SMA, and
LOW can populate this region in the ${\rm [NC]/[CC]} - A_{\rm N-D}$
plane. As we have discussed in Sec.~\ref{sec:r-a}, a detailed study 
of the zenith angle distribution of the charged current 
events during the night may discriminate among these solutions.

The zones of avoidance in the ${\rm [NC]/[CC]} - A_{\rm N-D}$ plane are:
all values of [NC]/[CC] larger than $5.2$ (for any values of $A_{\rm
N-D}$, [CC], and $\delta T$):  $A_{\rm N-D}$ less than $-0.02$ (for any
values of [NC]/[CC], [CC], and $\delta T$);  and [NC]/[CC] less than
$2.5$ together with  $A_{\rm N-D}$ larger than $0.02$.

\subsection{Correlation phenomenology}
\label{fig:correlncccvsa}

The correlations between [NC]/[CC] and $A_{\rm N-D}$ are similar to
the correlations between [CC] and $A_{\rm N-D}$ that were discussed in
Sec.~\ref{sec:r-a}.  The pointwise relation between [NC]/[CC] and
$A_{\rm N-D}^{\rm CC}$ for the LMA and LOW active solutions is washed out in
Fig.~\ref{fig:nca5} and Fig.~\ref{fig:nca8} by the fact that $A_{\rm
N-D}^{\rm CC}$ depends primarily on $\Delta m^2$ and [NC]/[CC]
primarily depends upon $\sin^2 2\theta$.

In all cases, the day-night effect is small for vacuum
oscillations. However, one can understand the general trend in
Fig.~\ref{fig:nca5} and Fig.~\ref{fig:nca8}, according to which the
solutions for VAC$_{\rm L}$ (larger $\Delta m^2$) correspond to smaller
values of [NC]/[CC] than the solutions for  VAC$_{\rm S}$. 

It is easy
to show from Eq.~(\ref{CC-NC}), Eq.~(\ref{P-vac}),  and  Eq.~(\ref{A-vac})
that for the  VAC$_{\rm S}$ solutions
\begin{equation}
\frac{\rm [NC]}{\rm [CC]}  = 1 + k \cdot A_{\rm N-D}~, 
\label{eq:linearncccvsa}
\end{equation}
where $k \equiv k (\Delta m^2)$ is a function of $\Delta m^2$ only.
For the VAC$_{\rm S}$ scenario, there are four islands of solutions along
which $\Delta m^2 \approx {\rm constant}$ but $\sin^2 2\theta$ changes.  So,
for a given VAC$_{\rm S}$ island $k \approx {\rm constant}$, where $k$
takes on a different value for each island of solutions. From
Eq.~(\ref{eq:linearncccvsa}), we see that [NC]/[CC] increases linearly
with $A_{\rm N-D}$. For $A_{\rm N-D} = 0$, we obtain [NC]/[CC] $= 1$.  All of
these features are apparent in Fig.~\ref{fig:nca5} and
Fig.~\ref{fig:nca8}.

The discussion following Eq.~(\ref{RNCCCst}) explains why for sterile
neutrinos [NC]/[CC] is predicted to be close to, but not identical to,
one. 

\section{[NC]/[CC] versus Shift of First Moment}
\label{sec:ncccdeltat}
Figures~\ref{fig:ncccm5} and \ref{fig:ncccm8} show, for electron
energy thresholds of $5$ MeV and $8$ MeV, maps onto the ${\rm[NC]/[CC]} -
\delta T$ plane of the currently-allowed regions in the $\Delta m^2 -
\sin^2 2\theta$ plane.  For illustrative purposes,
Fig.~\ref{fig:ncccm5} shows a simulated experimental point near the
current best-fit predicted point for the LMA solution. The error bars
are estimated $1\sigma$ uncertainties from Table~II of
Ref.~\cite{bks2000}, with a $1.3$\% fractional uncertainty in the
first moment and a $3.6$\% for the neutral to charged-current 
double ratio.

Figure~\ref{fig:ncccm5} and Fig.~\ref{fig:ncccm8} are similar to
Fig.~\ref{fig:ccm5} and Fig.~\ref{fig:ccm8}; the latter pair
refers to the correlation between the [CC] and the $\delta T$
variables.  Because of the relation Eq.~(\ref{R-NCCC}) between
[NC]/[CC] and [CC], the acceptable solution space, as well as the
zones of avoidance and the degree of overlap of the solution regions,
are similar in the two sets of figures.

\subsection{Discriminating among solutions}
\label{subsec:ncccvsdeltat}

\begin{figure}[!t]
\centerline{\psfig{figure=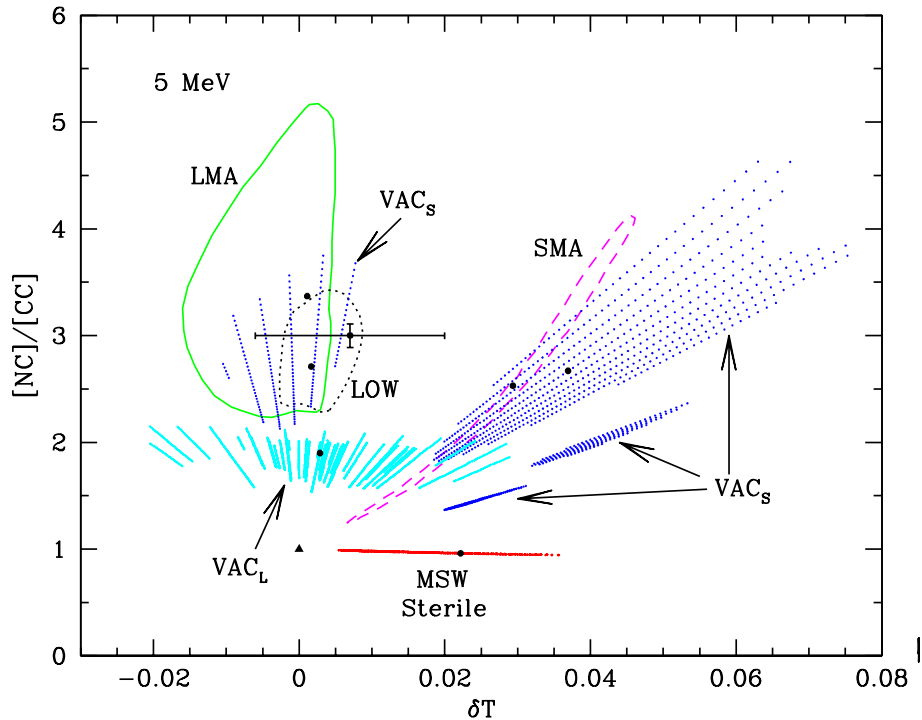,width=5.0in,angle=0}}
\tightenlines
\caption[]{\small The allowed regions for the the double ratio,
[NC]/[CC], versus the shift of the first moment, $\delta T$, for a
recoil electron energy threshold of $8$ MeV.  The meaning of the
symbols is the same as in Fig.~\ref{fig:cca5}, except that the regions
now refer to [NC]/[CC] and $\delta T$.
\label{fig:ncccm5}}
\end{figure}

There are certain regions of the parameter space, ${\rm [NC]/[CC]} -
\delta T$, that are relatively discriminatory. For example, for a $5$
MeV threshold (see Fig.~\ref{fig:ncccm5}), in the region in which $\delta
T > 0.03$ and $1 < {\rm [NC]/[CC]} < 5$, only the VAC$_{\rm S}$ and
the SMA solutions are represented. The most extreme values of the
[NC]/[CC] parameter, e.g., [NC]/[CC] $> 4.7 $ or $ < 1$, would
indicate, respectively, the LMA solution or the MSW Sterile solution.
Only VAC$_{\rm S}$ solutions have a fractional shift $\delta T >
4.5$\% .  Figure~\ref{fig:ncccm5} shows significant zones of
avoidance, e.g., [NC]/[CC] $> 2.2$ and $0.01 < \delta T < 0.01
{\rm [NC]/[CC]}$.

\begin{figure}[!ht]
\centerline{\psfig{figure=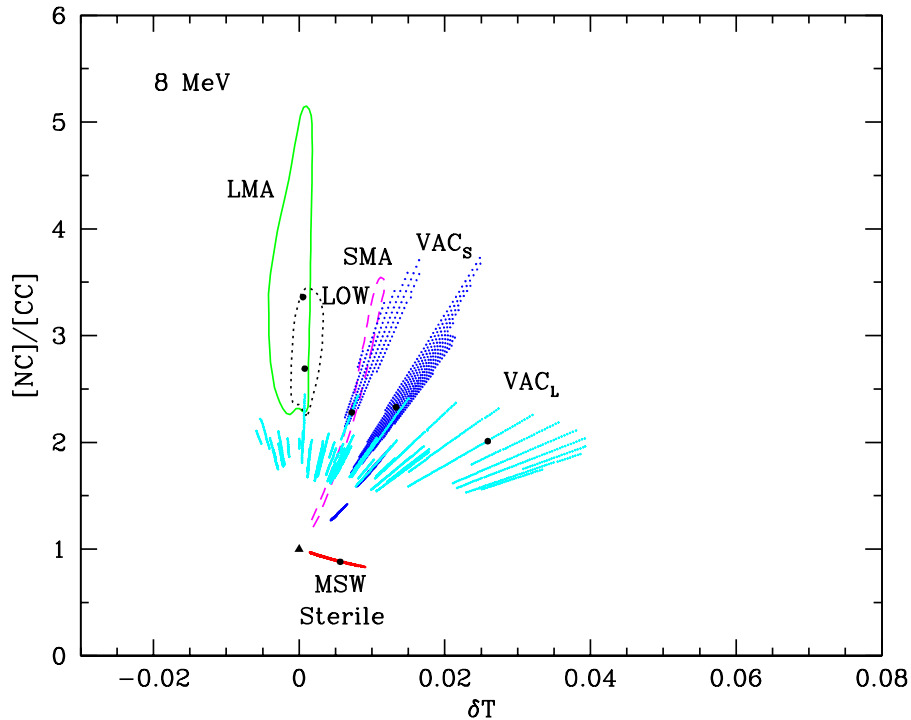,width=5.0in,angle=0}}
\tightenlines
\caption[]{\small The allowed regions for the the double ratio,
[NC]/[CC], versus the shift of the first moment, $\delta T$, for a
recoil electron energy threshold of $8$ MeV.  The meaning of the
symbols is the same as in Fig.~\ref{fig:cca5}, except that the allowed
regions refer to [NC]/[CC] and $\delta T$ and the threshold for the
recoil electron energy is $8$ MeV.
\label{fig:ncccm8}}
\end{figure}

For an electron recoil energy threshold of $8$ MeV  
(see Fig.~\ref{fig:ncccm8}), the degree of the overlap between the
predictions of different scenarios is not larger than it was for a $5$
MeV threshold, as is the case for ${\rm [CC]} - \delta T$ 
(cf. Fig.~\ref{fig:ccm5} and Fig.~\ref{fig:ccm8}). The allowed regions of 
VAC$_{\rm S}$ solution with largest $\Delta m^2$ do not overlap with
LMA and LOW regions, as occur for a $5$ MeV threshold. However, the 
largest $\Delta m^2$ VAC$_{\rm S}$ solutions do overlap with the
allowed SMA region for an $8$ MeV threshold.

\subsection{Correlation phenomenology}
\label{subsec:ncccvsdeltatphenom}

For the SMA solution, both [NC]/[CC] and $\delta T$ are determined by a
unique combination of neutrino variables, $\xi$, defined by
Eq.~(\ref{eq:xiparam}), so that 
the two measurables are strongly correlated: 
$\kappa_{\delta T-NC/CC}({\rm SMA}) \ll 1$.  
As the shift  $\delta T$ increases, the double ratio [NC]/[CC] 
also increases (cf. Fig.~\ref{fig:ncccm5} and  Fig.~\ref{fig:ncccm8}). 
Using Eq.~(\ref{CC-NC}) and Eq.~(\ref{LZ}) for the survival probability  and
Eq.~(\ref{mompar}) for the shift of the first moment, we find for an
electron threshold energy of $5$ MeV
\begin{equation}
\frac{\rm [NC]}{\rm [CC]} \approx e^{B \delta T}~, 
\label{NCCCmomcorr}
\end{equation}
where $B = 29.3$.

For the VAC$_{\rm S}$ solution, a strong correlation exists.    
Using Eq.~(\ref{CC-NC})
and Eqs.~(\ref{mom-vac}) and (\ref{P-vac}),   we find
\begin{equation}
\frac{\rm [NC]}{\rm [CC]} = 1 +  k'\delta T~, 
\label{T-RNCCC}
\end{equation} 
where $k' \equiv k'(\Delta m^2)$ is the function of the 
$\Delta m^2$ only. For a given VAC$_{\rm S}$ island,  $k'$ can be considered as a
constant. 
Thus the double ratio [NC]/[CC] is proportional to the shift $\delta T$.
The slope $ k'$ is different for different VAC$_{\rm S}$ 
islands.

For the MSW Sterile neutrino solution, the rate [NC]/[CC] is strongly
restricted by the measured value of ${\rm [ES]}_{\rm SK}$, whereas
$\delta T$ varies over a significant range  independent of [NC]/[CC].

The LMA solution does not predict a strong correlation, as can easily
be seen from Eqs.~(\ref{P-lma}) and (\ref{P-mom}).  The ratio 
[NC]/[CC] 
depends strongly on $\sin^2 2\theta$, whereas $\delta T$ depends
strongly on $\Delta m^2$.  The situation is similar for the LOW
solution.

\section{Conclusions}
\label{sec:sum}

We discuss and comment on in this section the principal results from
our analysis. We begin in Sec.~\ref{subsec:yesorno} with a restatement
of the problem we address and then describe in
Sec.~\ref{subsec:answer} our most important numerical results. In
Sec.~\ref{subsec:diagnosticpower}, we summarize how the predicted
correlations and zones of avoidance between neutrino measurables can
enhance the discriminatory power of solar neutrino experiments.
Finally, we discuss in Sec.~\ref{subsec:reducing} some additional work
that needs to be done in order to identify the correct set of
mixing angles and mass squared differences  of solar neutrinos.

\subsection{Correlated or not?}
\label{subsec:yesorno}

For a given pair of neutrino parameters, $\Delta m^2$ and $\sin^2
2\theta$, the predictions for all the neutrino observables (e.g., day-night
asymmetry or charged current rate) are completely determined. Thus on
a point-by-point basis the predictions for all the neutrino
measurables are fully correlated. But, the currently allowed
oscillation solutions constitute islands of finite size in the space
of neutrino parameters.

The practical question one wants to answer is:
For a specified set of allowed solutions (e.g., LMA or VAC$_{\rm
S}$), how well correlated are the predictions for different neutrino
measurables?  In other words, if one considers for example 
the predictions that
correspond to all the allowed values for $\Delta m^2$ and $\sin^2
2\theta$ currently included in the LMA solution, will the
predicted values of observables like the day-night 
asymmetry and the charged current
rate be strongly correlated? Or,  will  the range of 
$\Delta m^2$ and $\sin^22\theta$ within the allowed LMA domain obscure
the point-by-point correlations?

\subsection{The answer}
\label{subsec:answer}

Figures~\ref{fig:cca5} to \ref{fig:ncccm8} show the extent of the
predicted correlations between different neutrino observables in the
SNO experiment. These figures present our principal quantitative
results. The specifics of these correlations depend upon the data set
used (which will evolve with time) and the specified confidence level
($99$\% in this paper).

We have considered the following pairs of neutrino measurables:
$A_{\rm N-D}$ versus [CC] (day-night asymmetry for the charged
current, charged current rate), $\delta T$ versus [CC] (shift of the
first moment of the charged current electron recoil energy spectrum,
the charged current rate), $A_{\rm N-D}$ versus $\delta T$ (day-night
asymmetry, shift of first moment), [NC]/[CC] versus $A_{\rm N-D}$
(double ratio of neutral current to charged current rate, day-night
asymmetry), and [NC]/[CC] versus $\delta T$ (double ratio versus
shift of first moment).  For each pair of neutrino measurables, results are given
for two different electron recoil energy thresholds, $5$ MeV and $8$
MeV.  The correlations are discussed in the text following the figures
related to each pair of neutrino measurables.

Some of the currently-favored neutrino oscillation solutions
 predict strong correlations among measurable quantities. For example,
 the allowed set of SMA solutions predicts strong correlations between
 the values of  $A_{\rm N-D}$ and [CC], as well as between
 $\delta T$ and [CC] and between $A_{\rm N-D}$ and $\delta T$. On the
 other hand, the LMA solutions predict correlations only between
 $A_{\rm N-D}$ and $\delta T$ and not between $A_{\rm N-D}$ and [CC]
 or between $\delta T$ and [CC].
 
The correlations, and the lack of correlations, can be understood from
 simple analytic arguments. We derive in Sec.~\ref{subsec:examples}
 and in the Appendix approximate expressions giving the dependence of
 neutrino measurables upon $\Delta m^2$ and $\sin^2 2\theta$. In
 subsections labeled `correlation phenomenology,' we describe the
 physical bases for the correlations and for the lack of correlations
 between different pairs of neutrino observables.

\subsection{Diagnostic power: correlations and zones of avoidance}
\label{subsec:diagnosticpower}

Does the simultaneous analysis of different observables 
enhance the diagnostic power of solar neutrino
experiments? The answer is: ``Yes, in some cases.'' In subsections
labeled `Discriminating among solutions,' we emphasize for which cases
the correlations among the predictions are strongest and how they can
help in identifying the correct oscillation solution.  We give
examples in which multiple correlations can enhance the diagnostic
power. For example, the values predicted by the LMA oscillation
solution for the variables [NC]/[CC], $A_{\rm N-D}$, and $\delta T$
are all  correlated when $A_{\rm N-D}$ is large.  We also show by
examples that the dependence of the correlations on threshold provides
additional constraints on the allowed solar neutrino solutions.

The most powerful diagnostic pair that we have investigated may well
be [NC]/[CC] and  $A_{\rm N-D}$. Figures~\ref{fig:nca5} and
 \ref{fig:nca8}  display the results for this case. This pair is
particularly discriminatory because the systematic uncertainties in
[NC]/[CC] and  $A_{\rm N-D}$ can be reduced to values that are small
compared to the ranges of the observables that are shown in 
Figs.~\ref{fig:nca5} and \ref{fig:nca8}. Moreover, 
correlations are predicted between the values of [NC]/[CC] and
$A_{\rm N-D}$ for some favored oscillation solutions.  By contrast, 
correlations involving the charged current rate, [CC], are severely 
compromised by the uncertainty in the value of the neutrino absorption
cross section.  Figures~\ref{fig:ncccm5} and \ref{fig:ncccm8}
also show significant correlations between [NC]/[CC] and $\delta T$. 

Figures~\ref{fig:cca5} to \ref{fig:ncccm8} show that 
there are zones of avoidance in the parameter space of
 neutrino measurables. None of the currently favored neutrino
oscillation solutions predict values of the neutrino observables that
lie within these unoccupied regions. We identify some of the more
prominent zones of avoidance in the subsections `Discriminating among
solutions.' 

All of the currently favored oscillation solutions predict that the
zones of avoidance will not be populated by values from experimental
measurements. Thus an experimental test of whether or not the zones of
avoidance are populated by future measurements is a general test of
all of the presently allowed $2\nu$ oscillation solutions.

\subsection{Reducing the ambiguities}
\label{subsec:reducing}

Measurements with the SNO observatory will greatly reduce the allowed
regions in  neutrino parameter space. Will a unique solution emerge
from SNO measurements? Will we be able to identify the correct
oscillation solution as one of the six currently-favored islands?

There are many regions in Figs.~\ref{fig:cca5} to \ref{fig:ncccm8}
where multiple solutions (LMA, SMA, and LOW, e.g.) all overlap.  In
general, a unique identification will be possible only if one of the
variables lies near an extreme value in one of the observables planes
that we have considered in this paper. We will have to be somewhat
lucky to be able to extract a unique solution from SNO measurements alone.

In the future, we will study correlations between, on the one hand,
measurable quantities in the SNO and SuperKamiokande experiments, and,
on the other hand, quantities measured in low energy (less than 1 MeV)
solar neutrino experiments (such as BOREXINO~\cite{borexino}).  We
anticipate that unique inferences may be possible when low and high
energy solar neutrino measurements are combined.

Will the correlations and the zones of avoidance found in this paper
also be valid for more complicated schemes of neutrino mixing three or
even four neutrinos?  Extensive and detailed computations are
necessary in order to answer this question.
\acknowledgments

We are indebted to E. Kh. Akhmedov, E. Beier,
A. McDonald, and Y. Nir for valuable discussions 
JNB and AYS acknowledge partial support from NSF grant
No. PHY95-13835 to the Institute for Advanced Study and PIK
acknowledges support from NSF grant No. PHY95-13835 and NSF grant
No. PHY-9605140.

\appendix
\section{Dependence of observables on the oscillation
parameters}

In what follows we present simple expressions for solar neutrino observables
 which are valid in narrow intervals near the best-fit points for
 different oscillation solutions.  These expressions will be adequate
 for qualitative, and in many cases quantitative, understanding of the
 predicted correlations among the measurable quantities.  Details of
 the approximations and more precise expressions for the observables
 are given elsewhere \cite{NS}.

We use results of numerical calculations of the iso-contours 
 obtained in \cite{bkdn,bkl-mom} in order to find parameters in the   
analytical expressions. The allowed regions in the 
$\Delta m^2 -  \sin^2 2\theta$ plane are taken from
 Ref.~\cite{snoshow} (see Fig.~\ref{fig:global}). 

Before proceeding to the approximate expressions valid for different
oscillation solutions, we give the relevant definitions and equations
that were used in deriving the results presented in the following
subsections. 

For the MSW solution regions, 
the (daily) average  survival probability is given by 
(see Eq. (35) in \cite{NS}):   
\begin{equation}
P\equiv{1\over2}(P_D + P_N)
={1\over2}
\left[1 - \cos2\theta_S (1- 2P_1)(f_{reg} - \cos2\theta)\right], 
\label{barP}
\end{equation}
where $P_D$ and  $P_N$ are the averaged probabilities 
$P(\nu_e \rightarrow \nu_e)$ during the day time and during the night
time, respectively.

The quantities that appear in  Eq.~(\ref{barP}) are defined as follows. 

The probability   
$P_1 \equiv P(\nu_e \rightarrow \nu_1)$ is the probability that the
solar neutrinos reach the surface of the Earth as the mass eigenstate 
$\nu_1$.
  
The variable $\theta_S$ is the matter mixing angle in the neutrino production
region inside the Sun: 
\begin{equation}
\cos2\theta_S = {-1 + \eta_S \cos2\theta 
\over(1 - 2\eta_S \cos2\theta + \eta_S^2)^{1/2}}~,
\label{tSnusun}
\end{equation}
where 
\begin{equation}
\eta_S \equiv  {\Delta m^2/2 E V_S}~, 
\label{defetaS}
\end{equation} 
and $V_S$ is the matter potential in the center of the Sun;   

The regeneration factor, $f_{reg}$ (see Eq. (30, 32) in \cite{NS}),  
\begin{equation}
f_{reg} = {\eta_E\sin^22\theta\over2(1 - 2 \eta_E \cos2\theta +
\eta_E^2)},
\label{fregear}
\end{equation}
describes the Earth matter effect. The quantity $f_{reg}$ equals 
zero in absence of regeneration. In Eq. (\ref{fregear}),
\begin{equation}
\eta_E \equiv {\Delta m^2/2 E V_E}, 
\label{defetaE} 
\end{equation}
and $V_E$ is the effective  matter potential for the Earth.

The Day--Night asymmetry is given by (see Eq. (37) in \cite{NS}): 
\begin{equation}
A_{\rm N-D}\equiv{P_N - P_D \over P}
= {2 f_{reg} \over 1/(1 - 2 P_1) - \cos2\theta + f_{reg}}~, 
\label{defAND}
\end{equation}
where we have taken into account that in the 
$\Delta m^2$ region of significant 
Earth matter  effect:   
$\eta_S \ll 1$, and therefore  $\cos2\theta_S \approx -1$. \\

In calculating observables, the probability and asymmetry should be
averaged over the neutrino energy. The effect of averaging can be
represented by substituting for $2 E V$ the effective parameters $m^2$
which are introduced in different equations below. The values of the
effective $m^2$ should be determined from the results of exact
numerical calculations.\\

\subsection{LMA solution}

In the LMA solution region one has    
$\eta_E \gg 1$, so that according to 
(\ref{fregear}) the regeneration factor 
can be approximated by 
\begin{equation}
f_{reg} \approx {\sin^2 2\theta \over 2 \eta_E}~.     
\label{freglma}
\end{equation}
Moreover, in this region $P_1 \approx \cos^2 \theta$ 
and $\eta_S \ll 1$. Then expanding 
$\cos2\theta_S$ given in Eq. (\ref{tSnusun}) in powers of  $\eta_S$ 
and using the 
approximate expression Eq.~(\ref{freglma}) for $f_{reg}$ we obtain 
from Eq.~(\ref{barP}) the average survival probability  

\begin{equation}
P  \approx \sin^2 \theta + \frac{1}{4} \sin^2 2 \theta 
\left[\cos 2\theta  \left(\frac{\Delta m^2}{m_S^2}\right)^2 
+ \frac{m_E^2}{\Delta m^2} \right]~, 
\label{P-lma}
\end{equation} 
where $m_S^2$ and $m_E^2$ are fit parameters. 
The survival probability (and therefore 
the rate [CC] and the double ratio [NC]/[CC]) depend 
mainly on the mixing angle, $\sin^2 \theta$;   
 the dependence on $\Delta m^2$ is weak.  
The first term in the brackets of Eq.~(\ref{P-lma}) is   the correction
due to  effect of the adiabatic edge of the suppression pit,  
 which is due to closeness of the resonance to  the production point. 
This leads to deviation of a $P$ from $\sin^2 \theta$; 
$m_S^2$ is a parameter which corresponds to the product of the effective 
matter potential in the center of the sun, $V_S$, and the neutrino 
energy: $m_S^2 \sim 2\bar{E} V_{S}$. 
The second term in the brackets
describes the earth regeneration effect, where
$m_E^2 \sim 2\bar{E} V_{E}$.

From Eqs.~(\ref{firstmom}) and (\ref{P-lma}),  we find  for the shift of the
first moment  
\begin{equation}
\delta T  \approx 
\frac{\sin^2 2 \theta}{2(1 - \cos 2\theta)}
\left[ -2 \cos 2\theta \left(\frac{\Delta m^2}{m_S^2}\right)^2  +
\frac{m_E^2}{\Delta m^2} \right]~.
\label{P-mom}
\end{equation}
Here the  negative term in the brackets is due to  the adiabatic edge and 
the positive term describes the distortion due to the earth
regeneration effect.
The shift $\delta T$ is negative in the large-$\Delta m^2$ 
part of the LMA region and it becomes positive  
in the small $\Delta m^2$ part. 
For fixed $\Delta m^2$,  the shift  
$\delta T$ increases with decreasing $\sin^2 2 \theta$.  
For the reasons stated in the previous
paragraph, $\delta T$ is very small for the LMA solution.  

For the day-night asymmetry, we find from Eqs. (\ref{defAND}) 
and (\ref{freglma}) the following analytical result:   

\begin{equation}
A_{\rm N-D} \approx 
\frac{m_E^2}{\Delta m^2}
\left[\frac{1 - \cos 2\theta}{\sin^2 2\theta} + \frac{m_E^2}{2\Delta m^2} 
+ \frac{1}{2}\left(\frac{\Delta m^2}{m_S^2}\right)^2
\right]^{-1}~. 
\label{A-lma} 
\end{equation}
The asymmetry is, to a good approximation, inversely 
proportional to $\Delta m^2$. 
The first  term in brackets leads to 
a decrease of the asymmetry when the mixing  approaches maximal 
value.  The second term is due to the regeneration effect in the earth. 
The last term describes the effect of the adiabatic edge which 
becomes important for $\Delta m^2 \sim 10^{-4}$ eV$^2$. 
For smaller $\Delta m^2$ the latter  can be neglected and we obtain from 
Eq.~(\ref{A-lma}) the equations for the iso-asymmetry lines: 
\begin{equation}
\Delta m^2 \approx m_E^2 \left(\frac{1}{A_{\rm N-D}} - \frac{1}{2} \right)
\frac{\sin^2 \theta}{1 - \cos 2\theta}. 
\label{iso-lma}
\end{equation}

Comparing with results of numerical calculations,  we find 
$m_S^2 \approx 6 \times 10^{-5}$ eV$^2$ and 
$m_E^2 \approx 3 \times 10^{-6}$ eV$^2$.


\subsection{SMA solution}

  For not too small mixing angles,  
the survival probability  can be well described by the Landau-Zenner
formula \cite{LZ}:  
\begin{equation}
P_{\rm LZ}  \approx e^{- \frac{\xi}{\xi_0}},  
\label{LZ} 
\end{equation}
where $\xi \equiv \Delta m^2 \cdot \sin^2 2\theta $ and $\xi_0 (\sim 2
E/r_0)$ is a fit parameter, $r_0$ is the electron density scale
height, $n_e(r) \propto exp(-r/r_0)$.  The effect of earth
regeneration on the survival probability can be neglected here.  From
Eq.~(\ref{R-R}) we find the reduced rate
\begin{equation}
{\rm [CC]} \approx \frac{\rm [ES]_{\rm SK}}{1 - r + r e^{\xi/ \xi_0} }.
\label{r-sma}   
\end{equation}
 
The day-night asymmetry can be parametrized in the following way:  
\begin{equation}   
A_{\rm N-D} = \sin^2 2\theta \cdot f(\Delta m^2) \cdot 
\left[\frac{1 - 2 P_{\rm LZ} (\xi)}{P_{\rm LZ} (\xi)} \right]~, 
\label{A-sma}
\end{equation}
where 
\begin{equation}
f(\Delta m^2) \approx A \frac{(\Delta m^2/m_0^2)^3}{(\Delta m^2/m_0^2)^5 +
1},  
\label{ffunc}                          
\end{equation}
and $A = 7.8 $ and $m_0^2 = 3 \times 10^{-6}$ eV$^2$ are the fit
parameters.  Notice that $m_0^2$ corresponds to the $\Delta m^2$ with
which neutrinos with an average detected energy resonate in matter of
the earth.  The probability $P_{\rm LZ}(\xi)$ is given in Eq.~(\ref{LZ}).

For the shift of the first moment, Eqs.~(\ref{firstmom}) 
and  (\ref{LZ}) yield  
\begin{equation}
\delta T  =  \frac{B}{{E}} \Delta m^2 \cdot \sin^2 2 \theta   
= \frac{B}{{E}} \xi, 
\label{mompar}
\end{equation}
where $B$ is a fit parameter.

\subsection{LOW solution} 
In the LOW region,  the probability $P_1$ equals the jump probability 
and can be approximated by the generalized Landau-Zenner 
probability,   $P_1 \approx P_{LZ}'$. Also,  in the LOW region 
$\eta_E \ll 1$, and therefore the regeneration factor (\ref{fregear})
becomes 
\begin{equation}
f_{reg} \approx \sin^2 2\theta ~{\eta_E \over 2}~.
\label{freglow}
\end{equation}
Therefore the survival probability, Eq.~(\ref{barP}),  can be written using 
Eq.~(\ref{freglow}): 

\begin{equation}
P  \approx \sin^2 \theta + 
\frac{1}{4} \sin^2 2\theta \frac{\Delta m^2}{{(m_E^{\prime})^2} }
+ P_{\rm LZ}' \left(\cos 2\theta - \frac{\Delta m^2}{2{(m_E^{\prime})^2}} 
\sin^2 2\theta \right)~.  
\label{P-low}
\end{equation} 
where ${(m_E^{\prime})^2 \sim 2 E V_E}$.  Here the second term is the
correction due to the earth regeneration effect which is important for
the large $\Delta m^2$-part of the LOW region and the third term,
which is proportional to the jump probability, $P_{\rm LZ}'$ (defined
below), is due to effect of the non-adiabatic edge.  The generalized
Landau-Zenner probability, $P_{\rm LZ}'$, which is valid for large
vacuum mixing in the LOW region (see Ref.~\cite{pk}),  equals
\begin{equation}
P_{\rm LZ}' \approx e^{-\gamma \sin^2 \theta}~,   
\label{P-general}
\end{equation} 
where $\gamma =  2 \pi r_0 \Delta m^2 /2E$ and $r_0$ is the density
scale height of the solar electron density distribution.

From Eq.~(\ref{firstmom}) and Eq.~(\ref{P-low}), we find 
for the first moment  
\begin{equation}
\delta T  \approx 
- \frac{\sin^2 2\theta}{\sin^2 \theta} \frac{\Delta m^2}{4
{(m_E^{\prime})^2}} 
+ P_{\rm LZ}' \frac{\Delta m^2}{m_{na}^2}
\left(\cos 2\theta - \frac{\Delta m^2}{{(m_E^{\prime})^2}}\sin^2 2\theta   \right).  
\label{mom-low}
\end{equation}
Here $m_{na}^2 (\sim E/\pi r_0)$ is the fit parameter.  The first term
in Eq.~(\ref{mom-low}) gives the effect of the non-adiabatic edge and
the second term, which is proportional to $P_{\rm LZ}'$, describes the
distortion due to regeneration.  For fixed $\sin^2 2\theta$, the shift
decreases with increasing $\Delta m^2$. In the small $\Delta m^2$ part
of the allowed solution space, $\delta T$ is positive since the
spectrum is at the non-adiabatic edge of the suppression pit. The
shift is zero at $\Delta m^2 \approx 10^{-7}$ eV$^2$ and then becomes
negative due to the regeneration effect. The shift increases with
decreasing $\sin^2 2\theta$.

For the day-night asymmetry, we find 
\begin{equation}
A_{\rm N-D} \approx 
\frac{\Delta m^2}{{(m_E^{\prime})^2}}
\left[\frac{1 - \cos 2\theta}{\sin^2 2\theta}  + 
\frac{\Delta m^2}{2{(m_E^{\prime})^2}} + 
\frac{2P_{\rm LZ}'}{(1 - 2P_{\rm LZ}')\sin^2 2\theta}\right]^{-1} ~. 
\label{A-low} 
\end{equation}
The fit parameter ${(m_E^{\prime})^2} = 2.5 \times 10^{-6}$ eV$^2$. 
In the region of the LOW solutions, the  last term in brackets 
(containing $P_{\rm LZ}'$) describes 
the effect of adiabaticity breaking and is positive, which suppresses
the asymmetry.
The asymmetry  increases with $\Delta m^2$, 
in contrast with the behavior of the LMA solution.

\subsection{Vacuum oscillation solutions}  

The standard expression for the vacuum oscillation probability leads
to the following approximate relation: 
\begin{equation}
P = 1 -  \sin^2 2\theta \cdot \sin^2 \frac{\Delta m^2}{m^2_V}~.
\label{P-vac}
\end{equation}  
where $m^2_V~~ (\sim 4E/R)$ is a fit  parameter.  
We find for the shift of the first moment: 
\begin{equation} 
\delta T \sim R
\frac{1}{P} \frac{\Delta m^2}{m^2_0} \sin^2 2\theta \cdot 
\sin \left(\frac{2\Delta m^2}{m^2_V}\right)~. 
\label{mom-vac}
\end{equation}

The day-night asymmetry originates from 
the eccentricity of the earth's orbit and the existence of 
seasons \cite{bks2000}. 
We find the residual asymmetry (after 
 removal of the $R^{-2}$ dependence of the total flux) 
which is  related to the dependence of the oscillation probability on 
distance from the sun. The stronger the dependence of 
$P$ on the distance (oscillation phase)  the larger the asymmetry. 
Clearly,  the asymmetry is absent for $P = {\rm constant}$. 
Therefore
\begin{equation}
A_{\rm N-D} \propto  \frac{R}{P} \frac{dP}{dR} \propto  
\frac{1}{P} \frac{\Delta m^2}{m^2_V} \sin^2 2\theta \cdot
\sin 2\frac{\Delta m^2}{m^2_V}~. 
\label{A-vac}
\end{equation}
The  expression in Eq.~(\ref{A-vac}) coincides with 
$\delta T$, so $A_{\rm N-D}  \propto \delta T $.   This result was
obtained earlier (see the discussion following Eq.~(\ref{eq:A-mom1}) using
the fact that $P = P(R/E)$.

For the VAC$_{\rm L}$ solution,   
there are  strong averaging effects, 
and Eq.~(\ref{P-vac}) with a fixed characteristic energy 
does not reproduce accurately   
 the functional dependence of the survival probability on 
oscillation parameters. As a consequence, the relations   
Eq.~(\ref{mom-vac}) and Eq.~(\ref{A-vac}) describe only very
approximately the dependence upon neutrino
parameters of the ${\rm VAC_{\rm L}}$ solution.  

\subsection{MSW Sterile solution}

  The rate for the MSW Sterile solution is essentially fixed 
by the measured SuperKamiokande rate, ${\rm [ES]_{\rm SK}}$. 
The distortion of the electron recoil energy spectrum and the
day-night asymmetry
are similar to that for the SMA case, but the earth 
regeneration effect is much smaller.

\end{document}